\documentclass[aps,prd,preprint,floats,superscriptaddress,nofootinbib]{revtex4}
\usepackage[utf8]{inputenc}
\usepackage{amsmath,amssymb,bbm}
\usepackage{dsfont}
\usepackage{graphicx}
\usepackage{xcolor,slashed}
\usepackage[nottoc]{tocbibind}
\pdfoutput=1
\usepackage{appendix}
 \usepackage{hyperref}       
 \usepackage{url}            
 \usepackage{booktabs}       
 \usepackage{amsfonts}       
 \usepackage{nicefrac}       
 \usepackage{microtype}      
 \usepackage{lipsum}

\begin{document}

\title{Leptoquark-mediated Dirac neutrino mass and its impact on  $B \to K \nu \bar{\nu}$  and  $K \to \pi \nu \bar{\nu}$  decays}
\author{Chuan-Hung Chen}
\email[E-mail: ]{physchen@mail.ncku.edu.tw}
\affiliation{Department of Physics, National Cheng-Kung University, Tainan 70101, Taiwan}
\affiliation{Physics Division, National Center for Theoretical Sciences, Taipei 10617, Taiwan}

\author{Cheng-Wei Chiang}
\email[E-mail: ]{chengwei@phys.ntu.edu.tw}
\affiliation{Department of Physics and Center for Theoretical Physics, National Taiwan University, Taipei 10617, Taiwan}
\affiliation{Physics Division, National Center for Theoretical Sciences, Taipei 10617, Taiwan}

\author{Leon M.G. de la Vega }
\email[E-mail: ]{leonmgarcia@phys.ncts.ntu.edu.tw}
\affiliation{Physics Division, National Center for Theoretical Sciences, Taipei 10617, Taiwan}

\begin{abstract}

Right-handed neutrinos $\nu_R$ play a crucial role in flavor-changing neutral-current processes with missing energy, such as $b\to s + \slashed{E} $ and $d\to s + \slashed{E} $, where Belle-II reports unexpectedly large branching fraction in $B\to K \nu \bar\nu$ decays.  Assuming $\nu_R$ is the partner of the active neutrino $\nu_L$ in the standard model, a Dirac-type neutrino framework emerges.  We investigate a scenario of radiative Dirac neutrino mass generation in a scalar leptoquark (LQ) model with a global $U(1)_X$ symmetry to suppress Majorana mass, tree-level Dirac mass, and diquark couplings.  The simplest LQ realization consists of two $S_1 = (3,1,-1/3)$ LQs with distinct $U(1)_X$ charges.  A non-Casas-Ibarra parametrization is proposed to match neutrino data with fewer model parameters.  Imposing current experimental constraints from meson mixing and lepton flavor-violating processes, we find that right-handed neutrino effects can significantly enhance $B\to K^{(*)} \nu\bar{\nu}$ and $K^+\to \pi^+ \nu\bar{\nu}$.  Additionally, the model predicts excesses in $R_D$ from $B\to D\tau\bar{\nu}$ that remain within $1\sigma$ of current experimental data.

\end{abstract}
\maketitle


\section{Introduction}

Since the first discovery of neutrinos almost 70 years ago, there are still properties about them that we do not fully understand.  For example, as they are electrically neutral, the neutrinos are allowed to be of Majorana type or Dirac type.  The former scenario would violate lepton number conservation by two units, while the latter preserves it.  A clear distinction could be established through the observation of, for example, neutrinoless double beta decays.  However, such processes are highly suppressed because of the tiny neutrino mass.  Non-observable of such a signal motivates the exploration of the latter scenario by employing radiative mechanisms to generate Dirac-type masses and the studies of their implications on flavor physics.

Recently, Belle II reported evidence of the $B^+\to K^+ \nu\bar\nu$ decay~\cite{Belle-II:2023esi}.  Combining this result with earlier measurements from BaBar~\cite{BaBar:2013npw} and Belle~\cite{Belle:2013tnz,Belle:2017oht,Belle-II:2021rof}, the weighted average branching ratio (BR) is determined to be
 \begin{align}
  {\cal B}(B^+ \to K^+ \nu \bar\nu) =(1.3\pm 0.4)\times 10^{-5}\,.
 \end{align}
This measurement deviates from the standard model (SM) prediction, ${\cal B}(B^+\to K^+ \nu \bar\nu)^{\rm SM}=(4.92\pm 0.30) \times 10^{-6}$~\cite{Buras:2022qip}, by $2.7 \sigma$, suggesting a possible manifest of new physics in the $b\to s \slashed{E}$ transitions~\cite{Buras:2014fpa,Browder:2021hbl,Asadi:2023ucx,Athron:2023hmz,Bause:2023mfe,Allwicher:2023xba,Felkl:2023ayn,Wang:2023trd,Amhis:2023mpj,He:2023bnk,Chen:2023wpb,Datta:2023iln,Altmannshofer:2023hkn,Ho:2024cwk,Chen:2024jlj,He:2024iju,Hou:2024vyw,Buras:2024ewl,Kim:2024tsm,Hati:2024ppg,Buras:2024mnq,Hu:2024mgf,Zhang:2024hkn,Dev:2024tto,Calibbi:2025rpx,He:2025jfc,Bolton:2025fsq}.  Additionally, the theoretically clean decay $K^+\to \pi^+ \nu \bar \nu$ has been measured with a BR of ${\cal B}(K^+\to \pi^+ \nu \bar\nu)=(11.4^{+4.0}_{-3.3})\times 10^{-11}$~\cite{ParticleDataGroup:2024cfk}.  Though currently with large experimental errors, if a significant deviation from the SM prediction, ${\cal B}(K^+\to \pi^+ \nu \bar\nu)^{\rm SM}=(8.60\pm 0.42)\times 10^{-11}$~\cite{Buras:2022qip}, is confirmed, this channel would provide further evidence for new physics~\cite{Buras:2024ewl,Aebischer:2020mkv}.

In this work, we aim to construct a model that not only accommodates tiny Dirac-type neutrino mass but, more importantly, addresses the aforementioned intriguing processes involving $d_i \to d_j \nu \bar\nu$ transitions, where $d_{i,j}$ refer to $b,s,d$ quarks.  In particular, in addition to their role in generating Dirac neutrino masses, right-handed neutrinos are known to be able to significantly impact such processes involving missing energy in the final state~\cite{Browder:2021hbl,He:2021yoz,Rosauro-Alcaraz:2024mvx,Becirevic:2024iyi}.

On the other hand, leptoquarks (LQs), which mediate interactions between quarks and leptons, have recently garnered substantial attention in the literature due to their rich implications for flavor physics~\cite{Fajfer:2012jt,Sakaki:2013bfa,Calibbi:2015kma,Sahoo:2015qha,Bauer:2015knc,Chen:2017hir,Crivellin:2017zlb,Buttazzo:2017ixm,Gherardi:2020det,Gherardi:2020qhc,Crivellin:2019dwb,Davighi:2020qqa,Greljo:2021xmg,Crivellin:2021egp,Carvunis:2021dss,Davighi:2022qgb,Heeck:2022znj}.  Under the SM gauge symmetry, LQs can be classified as $SU(2)_L$ triplet, doublet, and singlet~\cite{Crivellin:2021egp}.  As a minimal extension, we will focus on the scalar $SU(2)_L$ singlet LQ, $S_1 = (3,1,-2/3)$, where the numbers in the parentheses represent the quantum numbers under $SU(3)_C\times SU(2)_L\times U(1)_Y$.  The $S_1$ framework is particularly compelling for two reasons: (i)  neutrino masses are generated radiatively, and (ii) its unique couplings lead to $d_i \to d_j \nu \bar\nu$ transitions at tree level, while the $d_i\to d_j \ell^- \ell^+$ decays at tree level are absent.

Without introducing additional symmetry, however, using the LQ as a mediator for Dirac neutrino masses presents two major challenges.  First, the Dirac mass term generated by the Brout-Englert-Higgs mechanism at tree level involves a Yukawa coupling for the heaviest neutrino mass at the order of $10^{-12}$, some eight orders of magnitude smaller than that of the $\tau$-lepton.  Second, LQ can couple to diquarks, potentially resulting in proton instability.  We find that introducing an additional $U(1)_X$ global symmetry can address these concerns, though it requires two $S_1$ LQs with distinct $U(1)_X$ charges.

If $U(1)_X$ is an exact symmetry across all sectors of the model, the Dirac neutrino mass cannot be generated radiatively because the symmetry also suppresses the corresponding mass term at tree level.  To circumvent this difficulty, a soft-breaking $U(1)_X$ term, which induces mixing between the two LQs, is introduced in the scalar potential.  Since the absence of this soft-breaking term would restore the symmetry, it can be regarded as a technical naturalness when a small parameter value is adopted to fit the tiny neutrino mass.  Interestingly, the soft-breaking term does not completely break $U(1)_X$.  With appropriate charge assignments, a residual $Z_3$ discrete symmetry remains, thereby suppressing diquark interactions to higher orders.

The introduction of flavor-dependent LQ couplings will inevitably add many new free parameters to the model.  To reduce such freedom, we propose a novel parametrization for the Yukawa couplings that is different from the Casas-Ibarra parametrization (CIP)~\cite{Casas:2001sr} which relies on a complex orthogonal matrix.  Our approach, based on the Dirac mass matrix and neutrino oscillation data, reduces the number of free parameters in the model, while significantly enhancing the BRs for both $B\to K \nu \bar\nu$ and $K\to \pi \nu \bar\nu$.

Moreover, we find that lepton-flavor violation (LFV) processes, such as loop-induced $\ell_i \to \ell_j \gamma$ and tree-level $\mu-e$ conversion in nuclei, impose the most stringent constraints on the relevant model parameters.  While a Casas-Ibarra-like parametrization suppresses semileptonic $B$ and $K$ decays mediated by LQ contributions, our parameterization allows right-handed neutrino currents to contribute to $d_i \to d_j \nu \bar\nu$ decays when left-handed neutrino contributions are suppressed by the LFV constraints, particularly those from the $\mu$-$e$ conversion.  Additionally, the ratio $R_D = {\cal B}(B \to D \tau \nu) / {\cal B}(B \to D \ell \nu)$ can be enhanced to be within the $1\sigma$ uncertainties of current experimental data.

The paper is organized as follows: In Sec.~\ref{sec:model}, we introduce the LQ model and present the detailed Yukawa couplings based on the new global $U(1)_X$ symmetry.  With these Yukawa couplings, we derive the one-loop-induced Dirac neutrino mass matrix in Sec.~\ref{sec:Dirac_Mass}.  To reduce the number of free parameters, we propose a new parametrization of Yukawa couplings, distinct from the conventional CIP approach.  Additionally, various simplified schemes under our new parametrization are discussed. The effective Hamiltonian relevant to the phenomenology studied in this work is presented in Sec.~\ref{sec:phe}.  In particular, we check that the model remains consistent with strict constraints from rare flavor-changing neutral currents (FCNCs) and LFV processes.  Detailed numerical analyses and discussions are presented in Sec.~\ref{sec:NA}.  Finally, we summarize our findings in Sec.~\ref{sec:summary}.

\section{Model and Yukawa couplings} \label{sec:model}

As a minimal extension of the SM, we introduce three right-handed neutrinos and two $S_1=(3,1,-2/3)$ LQs \footnote{This model for neutrino mass is similar to one classified as T1-i-1-A(C) in Ref.~\cite{Jana:2019mgj}.}. The tree-level Dirac mass term and the diquark couplings of $S_1$ can be suppressed by imposing a $U(1)_X$ global symmetry, under which the two $S_1$ are charged differently.  Since the effects of an additional $Z'$ gauge boson are irrelevant to the scope of this study, it suffices to keep the symmetry global to achieve our objective
\footnote{To protect the $U(1)_X$ symmetry, one may want to gauge it.  However, such a symmetry would be anomalous within the current model setup.  Of course, the gauge anomalies can be arranged to cancel by introducing additional matter contents, presumably at higher energy scales and irrelevant to our current study.}.
For clarity, the quantum numbers of the involved particles are given in Table~\ref{tab:charges}.  Here, $L$ ($Q_L$) refers to the $SU(2)_L$ lepton (quark) doublet; $(\ell_R, u_R, d_R)$ represent the right-handed lepton, up-type, and down-type quarks, respectively; $H$ is the SM Higgs doublet, and $S$ and $S'$ are the two LQs.  To avoid spontaneous symmetry breaking of $U(1)_X$, the SM Higgs does not carry the $U(1)_X$ charge.

\begin{table}[htp]
\caption{Quantum numbers of the involved particles under various symmetry groups.  We suppress the generation index of the fermions.}
    \label{tab:charges}
    \begin{center}
        \begin{tabular}{c|ccccc}
        \hline\hline
                & ~~$SU(3)_C$~~   & ~~$SU(2)_L$~~      & ~~$U(1)_Y$~~ & ~~$U(1)_{X}$ ~~ & ~~$Z_3$~~\\
                \hline
            ~~$L$~~         & 1              & 2           & $-1/2$      & $x$ &  $\omega$\\
            $\ell_R$         & 1              & 1           & $-1$       & $x$ & $\omega$\\
            $Q_L$         & 1              & 2           & 1/6      & $2x$ & $\omega^2$\\
            $u_R$         & 1              & 1           & 2/3      & $2x$ & $\omega^2$\\
            $d_R$         & 1              & 1           & $-1/3$     & $2x$ & $\omega^2$\\
              $\nu_R$         & 1              & 1           & 1     & $4x$ & $\omega$
              \\
            $H$             & 1 & 2 &1/2 &0  & 1 \\            
            $S$         & 3 & 1           & $-2/3$      & $3x$ & 1\\
                        $S'$         & 3 & 1           & $-2/3$      & $6x$  &1\\ 
         \hline\hline
 \end{tabular}
 \end{center}
        \end{table}

The scalar potential including the new colored-scalar LQs is given by
\begin{equation}
    \begin{split}
        V 
        =& 
        \mu^2_{H} H^\dagger H + \mu^2_{S}S^*S 
        + \mu^2_{S^\prime}S^{\prime*}S^\prime 
        + (\mu^2_{\text{mix}} S^* S^\prime + {\rm H.c.}) 
        + \lambda_1 (H^\dagger H )^2 +\lambda_2 (S^* S )^2  
        \\
        &+ \lambda_3 (S^{\prime*}S^\prime )^2 
        + \lambda_4 (H^\dagger H )(S^* S )
        + \lambda_5 (H^\dagger H )(S^{\prime*}S^\prime ) 
        + \lambda_6 (S^* S )(S^{\prime*}S^\prime ) 
        ~,
    \end{split}
    \label{eq:V}
\end{equation}
where the only non-self-Hermitian term is the soft-breaking term $S^* S'$.  In fact, such a term serves multiple purposes.  As stated earlier, if $U(1)_X$ were an exact symmetry, the Dirac mass term $\bar L \tilde {H} \nu_R$ would be forbidden to all orders of quantum corrections.  Therefore, the soft-breaking term allows the generation of Dirac neutrino mass radiatively.  Moreover, with the charge assignments given in Table~\ref{tab:charges}, the soft-breaking term preserves a residual $Z_3$ symmetry.  As a result, the diquark couplings and Majorana neutrino mass terms remain forbidden.

We parameterize the components of the Higgs doublet as
\begin{equation}
    H = \begin{pmatrix}
        G^+ \\
        \frac{1}{\sqrt{2}} (v + h + i G^0) 
    \end{pmatrix}\,,
\end{equation}
where $G^+$ and $G^0$ are the Goldstone bosons, $h$ is the SM Higgs boson, and  $v$ is the vacuum expectation value (VEV) of the Higgs field.  From Eq.~(\ref{eq:V}), the mass-squared matrix elements of $(S, S')$ are given by
\begin{equation}
\begin{split}
    M^2_{11}=&\mu^2_{S} + \frac{\lambda_4 v^2}{2} \,,\\
    M^2_{12}=&M^2_{21}=  \mu^2_{\text{mix}}  \,, \\
    M^2_{22}=& \mu^2_{S^\prime} + \frac{\lambda_5 v^2}{2} \,.
\end{split}
\end{equation}
With the assumption that $\mu^2_{\rm mix}$ is a real parameter, the mass-squared matrix can be diagonalized using a $2\times 2$ orthogonal matrix, parametrized as
\begin{equation}
\mathcal{O}_S
= \begin{pmatrix}
       \cos\theta_S & \sin\theta_S\\
       -\sin\theta_S & \cos\theta_S\\
   \end{pmatrix}\quad, \quad 
\end{equation}
where the mixing angle $\theta_S$ can be determined through the relation
 \begin{equation}
 \tan2\theta_S
 =
 \frac{2 \mu^2_{\text{mix}}}
 { \mu^2_S-\mu^2_{S^\prime} +\frac{1}{2}v^2(\lambda_4-\lambda_5) }\,.
 \end{equation}
To fit the small neutrino mass, it is required that $\mu^2_{\rm mix} \ll \mu^2_{S, S'}$, i.e., $\theta_S \ll 1$.  As a result, the mass eigenvalues of the physical LQ states $S_A$ and $S_B$ are approximately $m_{S_A}\simeq M_{11}$ and $m_{S_B}\simeq M_{22}$.  Accordingly, except for the mixing angle appearing in the neutrino mass matrix, the mixing effect between $S$ and $S'$ can be neglected when analyzing the LQ contributions to the FCNC processes.

Based on the charge assignments in Table~\ref{tab:charges}, the Yukawa couplings of LQs to quarks and leptons, consistent with the gauge symmetry, are given by
\begin{align}
    - \mathcal{L}_Y
    =&
     \bar{L} Y^\ell H \ell_R + \bar{Q}  Y^u \tilde{H} u_R  + \bar{Q} Y^d H d_R  + \overline{Q^C}\,  Y^{LL}\, i \tau_2\,  L\, S^*  
     \nonumber \\ 
     & +  \overline{u^C_R} \, Y^{RR}\, \ell_R S^{*}  + \overline{d^C_R} \, \overline{Y}^{RR}\,  \nu_R  S^{\prime *} + \text{H.c.} \,,
    \label{eq:Leptoquarkyukawas_0}
\end{align}
where flavor indices are suppressed, and $\tau_2$ denotes the second Pauli matrix.  We note that $S$ and $S'$ couple exclusively to left-handed and right-handed leptons, respectively.  The introduced soft-breaking effect in the scalar potential allows the right-handed neutrinos to couple with the left-handed neutrinos at the one-loop level.  Therefore, the Dirac neutrino mass can be radiatively generated.  Before considering radiative corrections to the neutrino mass, the mass matrices of quarks and charged leptons can be diagonalized by introducing the unitary matrices $V^u_{\chi}$, $V^d_\chi$, and $V^\ell_{\chi}$ with $\chi=R$ and $L$.  In terms of the physical states of quarks and charged leptons, the Yukawa couplings of the physical LQs can be expressed as
\begin{align}
    -\mathcal{L}_Y
    \supset&
     \left(   \overline{u^C_L}\,  \xi_1\,  \ell_L   +\overline{d^C_L}\, \xi_2\,  \nu_L +\overline{u^C_R}\,  \xi_4\, \ell_R   \right)  (\cos\theta_S S_A- \sin\theta_S S_B)
     \nonumber \\
     &+ \overline{d^C_R} \, \xi_3\, \nu_R (\sin\theta_S S_A+ \cos\theta_S S_B) 
     + \text{H.c. }
    \label{eq:Leptoquarkyukawas}
\end{align}
Here the matrices $\xi_i$ are defined as
\begin{equation}
    \xi_1= V^{u^T}_{L} Y^{LL} V^\ell_L\,,~~  
    \xi_2=-V^{d^T}_L Y^{LL} \,,~~ 
    \xi_3 = V^{d^T}_R \overline{Y}^{RR}\,,~~
    \xi_4= V^{u^T}_R\, Y^{RR} \, V^\ell_R\,.
    \label{eq:redefinedYukawas}
\end{equation}
Since the LQ couplings to the physical quark and lepton states are through the $\xi_i$ matrices, our subsequent analysis will focus on them rather than the Yukawa couplings of $Y^{LL}$, $Y^{RR}$,  $\bar{Y}^{RR}$, and the flavor mixing matrices $V^{u,d,\ell}_\chi$.  Using the Cabibbo-Kobayashi-Maskawa (CKM) matrix, $V_{\rm CKM} = V^{u^\dagger}_L V^{d}_L$, we obtain a relation between $\xi_1$ and $\xi_2$: $\xi_1 = -V^*_{\rm CKM} \xi_2 V^\ell_L$.  Since $\overline{ Y}^{RR}$ and $Y^{RR}$ in $\xi_{3,4}$ are independent, there is no simple relationship between them.  Based on the couplings shown in Eq.~(\ref{eq:Leptoquarkyukawas}), the one-loop-induced Dirac neutrino mass matrix will be related to $\xi_2$ and $\xi_3$.

\section{Radiative corrections to the Dirac neutrino mass and parametrization of Yukawa couplings} \label{sec:Dirac_Mass}

We discuss in this section how to generate the Dirac-type neutrino mass at the one-loop level in the model and show how the Yukawa couplings $\xi_2$ and $\xi_3$ are parametrized in a way that is different from the Casas-Ibarra parametrization.

\subsection{One-loop-induced Dirac neutrino mass}

\begin{figure}
    \centering
    \includegraphics[scale=1.]{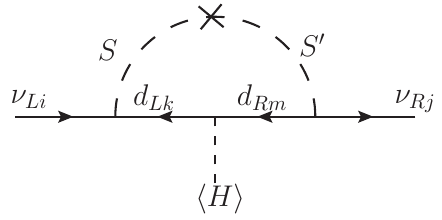}
    \caption{Feynman diagram for the Dirac neutrino mass generation mechanism with $S_1$ LQs. }
    \label{fig:neutrinomassfeynman}
\end{figure}

From the Yukawa couplings in Eq.~(\ref{eq:Leptoquarkyukawas}), the neutrinos interact with LQs and down-type quarks.  As a result, the Dirac neutrino mass matrix can be generated at the one-loop level, with LQs and down-type quarks serving as the mediators.  A representative Feynman diagram is depicted in Fig.~\ref{fig:neutrinomassfeynman}, where the cross represents the mixing effect, $\mu^2_{\rm mix}$, between $S$ and $S'$.  The $\langle H\rangle$ in the diagram is the Higgs VEV and indicates a mass insertion in the down-type quark propagator; therefore, the Dirac neutrino mass matrix is proportional to the down-type quark mass running in the loop.

Assuming $m_{S_B}>m_{S_A}$, the resulting neutrino mass matrix is given by:
 \begin{equation}
 {\bf m}^{\rm Dirac}_{\nu} 
 = 
 \frac{c_\theta s_\theta}{16 \pi^2}\ln\left(\frac{m^2_B}{m^2_A} \right) 
 \left( \xi^\dagger_2 {\bf m}_d \xi_3\right)\,, 
 \label{eq:Dirac_mass}
 \end{equation}
where $c_{\theta_S}=\cos\theta_S$, $s_{\theta_S}=\sin\theta_S$, and ${\bf m}_{d}$ denotes the diagonal down-type quark mass matrix.  The condition $m_{d_i}\ll m_{S_{A,B}}$ has been applied to obtain Eq.~\eqref{eq:Dirac_mass}. To diagonalize ${\bf m}^{\rm Dirac}_{\nu}$, unitary matrices $V^\nu_\chi$ are introduced as follows:
 \begin{equation}
 {\bf m}^{\rm dia}_{\nu} 
 = 
 V^{\nu^\dagger}_{L} {\bf m}^{\rm Dirac}_\nu V^\nu_R\,. 
  \label{eq:mD_dia}
 \end{equation}
As a result, the effects of $V^\nu_\chi$ must be incorporated into Eq.~(\ref{eq:Leptoquarkyukawas}). Specifically, the parameters $\xi_{2,3}$ are replaced by:
 \begin{equation}
 \xi_{2}\to \tilde{\xi}_2=\xi_2 V^\nu_L\,,~~ 
 \xi_3\to \tilde{\xi}_3= \xi_3 V^\nu_R
 \label{eq:xi_prime}
 \end{equation}
for the couplings to the massive neutrino states.  Furthermore, when the physical states are used to describe the weak charged-lepton current interactions, the Pontecorvo-Maki-Nakagawa-Sakata (PMNS) matrix is defined via the interaction term $\overline\ell \gamma^\mu V_{\rm PMNS} P_L \nu_\ell W^-_\mu$, where $V_{\rm PMNS}=V^{\ell^\dag}_L V^\nu_L$.

\subsection{Parametrization of $\xi_2$ and $\xi_3$}

According to Eq.~(\ref{eq:Dirac_mass}), the Yukawa couplings involved in ${\bf m}^{\rm Dirac}_\nu$ are $3\times 3$ complex matrices $\xi_{2,3}$.  In general, these matrices have independent free parameters that cannot be fully constrained by the neutrino oscillation data, which include measurements of three mass-squared differences, three mixing angles, and one CP-violating phase.  For our model to be more predictive, we consider the following different parametrization schemes for $\xi_{2,3}$ to reduce the number of free parameters.

In the scenario of Majorana neutrinos, it is known that the relevant Yukawa couplings can be linked to the neutrino masses, the PMNS matrix, and an undetermined complex orthogonal matrix through the CIP~\cite{Casas:2001sr}.  If the neutrino masses and PMNS matrix are treated as known observables, the remaining free parameters are associated with the complex orthogonal matrix.  We find that a Casas-Ibarra-like parameterization for Dirac-type neutrino mass in our model can be derived as follows.  Reformulating Eqs.~(\ref{eq:Dirac_mass}) and (\ref{eq:mD_dia}), we obtain
  \begin{equation}
   {\bf m}^{\rm Dirac}_\nu 
   = 
   V^\nu_L {\bf m}^{\rm dia}_{\nu} V^{\nu^\dag}_R 
   = 
   \xi^\dag_2  \widetilde{\bf m}_d  \xi_3\,,
   \label{eq:Dm_para}
  \end{equation}
where $\widetilde{{\bf m}}_d = c_{\theta_S} s_{\theta_S} \ln(m^2_{S_B}/m^2_{S_A}) {\bf m}_d/(16\pi^2)$.  By introducing a unitary matrix $U$, the last equality in Eq.~(\ref{eq:Dm_para}) can be satisfied if $\xi_2$ and $\xi_3$ are parametrized as
 \begin{align}
 \begin{split}
 \xi_2 &= \frac{1}{\sqrt{\widetilde{\bf m}_d}}  U  \sqrt{{\bf m}^{\rm dia}_\nu}  V^{\nu^\dag}_L = \tilde{\xi}_2 V^{\nu^\dag}_L \,,  \\
 \xi_3 & =  \frac{1}{\sqrt{\widetilde{\bf m}_d}}  U  \sqrt{{\bf m}^{\rm dia}_\nu} V^{\nu^\dag}_R = \tilde{\xi}_3 V^{\nu^\dag}_R\,.
 \end{split}
 \label{eq:para}
 \end{align}
This parametrization is not unique, and the unitary matrix $U$ is independent of flavor mixing.  Its primary advantage is to ensure that ${\bf m}^{\rm Dirac}_\nu$ can be diagonalized.  Although Eq.~(\ref{eq:para}) involves three unknown matrices, $V^\nu_{L,R}$ and $U$ and the parametrization does not seem to reduce the number of parameters, the couplings relevant to physical processes, as shown in Eq.~(\ref{eq:xi_prime}), are $\tilde{\xi}_{2}$ and $\tilde{\xi}_{3}$ that depend only on the single unknown matrix $U$.  Hence, $\tilde{\xi}_2=\tilde{\xi}_3$ by the parametrization.  Once $\widetilde{\bf m}_d$ and ${\bf m}^{\rm dia}_\nu$ are fixed, the free parameters are only contained in $U$.  Using this relation, $\xi_1$ can be expressed as:
  \begin{equation}
  \xi_1=- V^*_{\rm CKM} \tilde{\xi}_2 V^\dag_{\rm PMNS}\,. \label{eq:xi_1}
  \end{equation}

Applying the restricted parametrization in Eq.~(\ref{eq:para}) to the phenomenological analysis reveals a key limitation: the BRs for the decays $B \to K \nu \bar{\nu}$ and $K \to \pi \nu \bar{\nu}$ cannot be simultaneously enhanced while satisfying the neutrino data and the experimental upper limits on the LFV processes.  Thus, the equality of $\tilde{\xi_2}=\tilde{\xi_3}$ needs to be violated. To achieve desired phenomenological purposes, we propose a new parametrization scheme:
 \begin{align}
 \begin{split}
 \xi_2 &= \frac{1}{\sqrt{\widetilde{\bf m}_d}} \eta^\dag U \epsilon^\dag \sqrt{{\bf m}^{\rm dia}_\nu}  V^{\nu^\dag}_L = \tilde{\xi}_2 V^{\nu^\dag}_L \,,  \\
 \xi_3 & =  \frac{1}{\sqrt{\widetilde{\bf m}_d}}  \eta^{-1} U \epsilon^{-1} \sqrt{{\bf m}^{\rm dia}_\nu} V^{\nu^\dag}_R = \tilde{\xi}_3 V^{\nu^\dag}_R\,,
 \end{split}
 \label{eq:para_1}
 \end{align} 
where $\epsilon$ is a non-singular matrix and its simplest patterns are:
\begin{align}
\begin{split}
    \epsilon_a &= \begin{pmatrix}
        \varepsilon_1 & 0 & 0\\
        0 & \varepsilon_2 & 0\\
        0 & 0  & \varepsilon_3 
    \end{pmatrix}\,,~\epsilon_b = \begin{pmatrix}
        0 & 0 & \varepsilon_1\\
        0 & \varepsilon_2 & 0\\
        \varepsilon_3 & 0& 0 
    \end{pmatrix}\,,~ \epsilon_c= \begin{pmatrix}
        \varepsilon_1 &  0 & 0\\
       0  & 0 &\varepsilon_2\\
        0 & \varepsilon_3 & 0
    \end{pmatrix}\,, \\
    \epsilon_d &= \begin{pmatrix}
        0&  \varepsilon_1 & 0\\
        \varepsilon_2  & 0 &0\\
        0 & 0 & \varepsilon_3
    \end{pmatrix}\,,~
     \epsilon_e  = \begin{pmatrix}
       0 &  \varepsilon_1 & 0\\
        0 & 0 &\varepsilon_2\\
        \varepsilon_3 & 0 & 0 
    \end{pmatrix}\,,~ \epsilon_f = \begin{pmatrix}
        0& 0 &  \varepsilon_1\\
        \varepsilon_2  & 0 &0\\
        0 & \varepsilon_3 & 0
    \end{pmatrix}\,.
    \end{split}
    \label{eq:e_para}
\end{align}
Here, $\varepsilon_k$ are generally free complex parameters.  The matrix $\eta$ can be any of the $\epsilon_i$ shown in Eq.~(\ref{eq:e_para}) with $\varepsilon_k=1$ for all $k$.  If $\varepsilon_k$ are real parameters, it can be shown that the nonzero matrix elements in $(\epsilon^{-1}_i)^\dag$ are located at the same entries as those in $\epsilon_i$, with the corresponding values replaced by $\varepsilon^{-1}_k$.  Theoretically, there is no way to determine which matrix patterns for $\epsilon$ and $\eta$ are most suitable for phenomenological purposes.  We will show that a preference is singled out by experimental observations.

\section{Phenomenology} \label{sec:phe}

In this section, we first formulate the effective Hamiltonian based on the Yukawa couplings in Eq.~(\ref{eq:Leptoquarkyukawas}) to describe the relevant processes, including $d_i \to d_j \nu \bar\nu$, $\ell_\alpha \to \ell_\beta \gamma$, $\ell_\alpha \to 3 \ell_\beta$, $\mu-e$ conversion at tree level, $\Delta F=2$ transitions (where $F$ denotes a neutral meson), and $b\to c (u) \tau \bar\nu$.  For the non-perturbative transition matrix elements appearing in $\mu-e$ conversion within nuclei, the mixing parameter of $F$ and $\bar F$, and the BRs for $B\to D^{(*)} \ell \bar\nu$, we employ the Python package Flavio~\cite{Straub:2018kue}, which uses the nuclear form factors in Ref.~\cite{Kitano:2002mt} for the $\mu-e$ conversion, the matrix elements from Ref.~\cite{Beneke:1998sy,Artuso:2015swg,FermilabLattice:2016ipl} for $F \to \bar F$ transitions, and the form factors in Ref.~\cite{Gubernari:2018wyi} for the $B\to D^{(*)}$ decays.  Additionally, we implement additional effective Hamiltonian into Flavio to estimate the BRs for processes such as $\ell_\alpha\to \ell_\beta \gamma$, $\ell_\alpha\to 3\ell_\beta$, and $B\to \tau \bar\nu$.  This integrated approach ensures a theoretically consistent numerical analysis.

\subsection{$B\to K^{*} \nu \bar\nu$ and $K\to \pi \nu \bar\nu$}

From the Yukawa couplings in Eq.~(\ref{eq:Leptoquarkyukawas}) and neglecting the small $\theta_S$ effects, the effective Lagrangian for $d \to d' \nu \bar\nu’$ transitions mediated by $S_A$ and $S_B$ at tree level can be expressed as:
 \begin{align}
 {\cal L}_{d \to d' \nu \bar\nu} 
 =& 
 \frac{1}{2 m^2_{S_A}}  (\tilde{\xi}_2)_{ij} (\tilde{\xi}_2)^\dag_{k \ell}\; \overline{d_L}_\ell \gamma_\mu d_{ Li} \; \overline{\nu_L}_k \gamma^\mu \nu_{L j}
 \nonumber \\ 
 & + \frac{1}{2 m^2_{S_B}} (\tilde{\xi}_3)_{i j} (\tilde{\xi_3})^\dag_{k \ell}\;  \overline{d_R}_\ell \gamma_\mu d_{ R i} \; \overline{\nu_R}_k \gamma^\mu \nu_{R j} \,. 
 \label{eq:Ld2nu}
 \end{align}
Since the quark and lepton currents in Eq~(\ref{eq:Ld2nu}) are vectorial and carry the same chirality, as in the corresponding SM currents, the resulting BR can be readily scaled up from the SM BR by an appropriate multiplicative factor.  Explicitly, the BR for $B\to K \nu \bar \nu$ can be formulated as~\cite{Browder:2021hbl}
 \begin{align}
 \begin{split}
  {\cal B} (B\to K \nu \bar\nu) 
  &= 
  \frac{{\cal B}(B\to K \nu \bar\nu)^{\rm SM}}{3}
   \left[ 
   \sum_{i } \left|1+ \frac{C^{ii }_{LL}}{ X^{\rm SM}_{LL}} \right|^2 
   + \sum_{i\neq j}  \left| \frac{C^{ij}_{LL}}{X^{\rm SM}_{LL}}\right|^2 
   + \sum_{i,j} \left| \frac{C^{ij}_{RR} }{X^{\rm SM}_{LL} }\right|^2
   \right]\,,
 \end{split}
 \end{align}
where $X^{\rm SM}_{LL}$, $C^{ij}_{LL}$, and $C^{ij}_{RR}$ are defined as:
 \begin{align}
 \begin{split}
  X^{\rm SM}_{LL} & = V_{tb} V^*_{ts} X_t \,, \\
  C^{ij}_{LL} & = - \frac{1}{C_{\rm SM}} \frac{1}{2 m^2_{S_A}}(\tilde{\xi}_2)_{3 j} (\tilde{\xi_2})^\dag_{ i 2}\,, \\
  C^{ij}_{RR} & = - \frac{1}{C_{\rm SM}} \frac{1}{2 m^2_{S_B}}(\tilde{\xi}_3)_{3 j} (\tilde{\xi_3})^\dag_{i 2}\,.
   \label{eq:Cij_B}
 \end{split}
 \end{align}
Here, $C_{\rm SM}=\sqrt{2} G_F \alpha/(\pi s^2_W)$, $s_W=\sin \theta_W$, $G_F$ is the Fermi decay constant, $V_{ij}$ are the CKM matrix elements, $\alpha=e^2/(4\pi)$, and $X_t \approx 1.481$~\cite{Buras:2015qea}.  In this model, the hadronic transition form factors only involve those in the SM.  Notably, the same applies to the $B\to K^* \nu \bar\nu$ decay. Therefore, to quantity the new physics effects, we define a ratio:
 \begin{equation}
 R^{\nu\nu}_K
 =
 \frac{{\cal B}(B\to K^* \nu \bar \nu)}
      {{\cal B}(B\to K^* \nu \bar \nu)^{\rm SM}} 
 =
 \frac{{\cal B}(B\to K\nu \bar \nu)}
      {{\cal B}(B\to K\nu \bar \nu)^{\rm SM}} \,.
 \end{equation}

Using Eq.~(\ref{eq:Ld2nu}) and the results in Refs.~\cite{Mescia:2007kn,Buras:2015qea,He:2018uey}, the BRs of the $K^+\to \pi^+ \nu \bar\nu$ and $K_L\to \pi^0 \nu \bar\nu$ decays can be obtained respectively as:
 \begin{align}
\begin{split}
{\cal B}(K^+ \to \pi^+ \nu \bar\nu) 
 \approx & \frac{ \kappa_+}{3}  \left\{ \sum_i \left[ \left( \frac{{\rm Im} X^{ii}_{LL} }{\lambda^5} \right)^2 
+ \left( \frac{{\rm Re}(V^*_{cs} V_{cd} )}{\lambda} P_c (X)+ \frac{{\rm Re}(X^{ii}_{LL})}{\lambda^5}\right)^2  \right] \right.  \\
&\left.  +  \sum_{i\neq j} \left| \frac{X^{ij}_{LL}}{\lambda^5} \right|^2 +  \sum_{i,j} \left| \frac{X^{ij}_{RR}}{\lambda^5} \right|^2
\right\}
~, \\
{\cal B}(K_L \to \pi^0 \nu \bar\nu) 
 \approx &  \frac{\kappa_L}{3} \left[ \sum_i \left( \frac{ {\rm Im}(X^{ii}_{LL}) }{ \lambda^5 }\right)^2 + \sum_{i\neq j} \left| \frac{X^{ij}_{LL} - X^{ji*}_{LL}}{2 \lambda^5} \right|^2 \right]  \\
 & +  \frac{\kappa_L}{3}  \sum_{i,j} \left|  \frac{X^{ij}_{RR} -X^{ji*}_{RR}}{2 \lambda^5}\right|^2 ~, 
\end{split}
\label{eq:KpL}
\end{align}
where $\lambda=|V_{us}|$, $\kappa_{+}=(5.173\pm 0.025)\times 10^{-11} (\lambda/0.225)^8$, $\kappa_L =(2.231\pm 0.013)\times 10^{-10}(\lambda/0.225)^8$, and $P_c(X)=0.405\pm 0.024$ denotes the charm-quark loop contributions.  The $X^{ij}_{LL}$ and $X^{ij}_{RR}$ factors are defined as:
\begin{equation}
\begin{split}
X^{ij}_{LL} = & V_{td} V^*_{ts}  X_t \delta_{ij} - \frac{1}{C_{\rm SM}} \frac{1}{2m^2_{S_A}} (\tilde{\xi}_2)_{1j } (\tilde{\xi_2})^\dag_{i 2}~, \\
X^{ij}_{RR} = &  - \frac{1}{C_{\rm SM}} \frac{1}{2m^2_{S_B}} (\tilde{\xi}_3)_{1j } (\tilde{\xi_3})^\dag_{i 2}~.
\label{eq:Xij_K}
\end{split}
\end{equation}
It is worth noting that $\tilde{\xi}_2$ is related to $\xi_1$, which is stringently constrained by the LFV processes.  As a result, the right-handed couplings associated with $\tilde{\xi}_3$ dominate the $d_i \to d_j \nu \bar\nu$ transitions.  Similar to the $B\to K \nu \bar\nu$ decay, we define the ratio:
  \begin{equation}
 R^{\nu\nu}_\pi=\frac{{\cal B}(K\to \pi \nu \bar \nu)}{{\cal B}(K\to \pi \nu \bar \nu)^{\rm SM}} \,,
 \end{equation}
where $\pi=\pi^+ (\pi^0)$ for the $K^+ (K_L)$ decay.

\subsection{$\ell_{\alpha} \to \ell_{\beta} \gamma$}

The radiative LFV processes arise from photon-penguin diagrams, where the photon is emitted from the LQ and the up-type quark in the loop.  Based on the Yukawa couplings in Eq.~(\ref{eq:Leptoquarkyukawas}), the effective Hamiltonian for $\ell_\alpha \to \ell_\beta \gamma$ is expressed as:
\begin{align}
{\cal H}_{\ell_\alpha \to \ell_\beta \gamma} & = \frac{e}{m_{S_A}} \bar \ell_\beta i \sigma^{\mu\nu} \left( B_L P_L + B_R P_R \right) \ell_\alpha \epsilon_\mu k_\nu \,, 
\label{eq:H_RLFV}\\
\end{align}
with
\begin{align}
\left( B_L \right)_{\beta \alpha}
&= 
\frac{m_t}{ 16 \pi^2 m_{S_A}}\left(Q_{S_1}I_1(r^t_A)+Q_u I_2(r^t_A) \right) (\xi^\dagger_4)_{\beta 3} (\xi_1)_{3\alpha} \,, \nonumber \\
\left( B_R \right)_{\beta \alpha}
&=
\frac{ m_t}{ 16 \pi^2 m_{S_A}}\left(Q_{S_1}I_1(r^t_A)+Q_u I_2(r^t_A) \right) (\xi^\dagger_1)_{\beta 3} (\xi_4)_{3\alpha} \,,
\end{align}
where $c_{\theta_S}\approx 1$ is applied, the small $s_{\theta_S}$ effect is neglected, $Q_{S_1}$ and $Q_u$ are the electric charges of $S_1$ and up-type quark, respectively, $r^t_A=m^2_t/m^2_{S_A}$, and the loop integrals $I_1$ and $I_2$ are defined as:
 \begin{equation}
 \begin{split}
I_1(r) & = -\frac{5-22 r + 5 r^2}{36 (1-r)^3} - \frac{(1-3r) \ln r}{6 (1-r)^4}\,, 
\\
I_2(r) & =\frac{-3 + r}{2 (1-r)^2} - \frac{\ln r}{(1-r)^3} \,.
 \end{split}
 \end{equation}
Since $\theta_S \ll 1$ and $m_{\ell_\beta}\ll m_{\ell_\alpha} \ll m_{S_A}$, lepton mass-dependent factors have been neglected.

If the initial- and final-state lepton flavors in Eq.~(\ref{eq:H_RLFV}) are identical, we can use it to evaluate the lepton's anomalous magnetic dipole moment ($g-2$) and electric dipole moment (EDM) as, respectively:
\begin{align}
 a_\ell 
 & = 
 \frac{m_\ell m_t}{8 \pi^2 m^2_{S_A}} 
 \left[ Q_{S_1}I_1(r^t_A)+Q_u I_2(r^t_A) \right] 
 {\rm Re}\left((\xi^\dagger_4)_{\ell 3} (\xi_1)_{3\ell}\right)\,, \\
 d_\ell & = \frac{e  m_t}{16 \pi^2 m^2_{S_A}} 
 \left[ Q_{S_1}I_1(r^t_A)+Q_u I_2(r^t_A) \right] 
 {\rm Im}\left((\xi^\dagger_4)_{\ell 3} (\xi_1)_{3\ell}\right)\,.
 \end{align}
As stated earlier, $\xi_1$ is related to $\tilde{\xi}_2$.  Therefore, the CP-violating phase, which contributes to the neutrino mass matrix, can have a significant impact on the lepton EDM.

\subsection{$\ell_\alpha \to 3 \ell_\beta$}

The LFV processes $\ell_\alpha \to 3 \ell_\beta$ in the model arise from photon- and $Z$-penguin diagrams and from box diagrams mediated by $S_A$. Although the $\ell_\alpha\to 3 \ell_\beta$ transitions from the penguin diagrams involve tensor-type currents (e.g., $\bar\ell_\beta i \sigma^{\mu \nu} \ell_\alpha k_\nu$), their contributions are negligible compared to the vector-type currents because $k_\nu \sim O(m_{\ell_\alpha})\ll m_{S_{A}}$. Thus, 
the effective Hamiltonian for $\ell_\alpha\to 3 \ell_\beta$ from the photon-penguin diagrams is
\begin{equation}
{\cal H}^\gamma_{\ell_\alpha \to 3 \ell_\beta} 
= 
2 \sqrt{2} G_F 
\left[ \bar u_\beta \gamma_\mu 
\left(C^\gamma_L P_L + C^\gamma_R P_R \right) 
u_\alpha \right]
\left[ \bar u_\beta \gamma^\mu v_\beta \right] \,,
\end{equation}
where the effective Wilson coefficients are given by
 \begin{equation}
 \begin{split}
 \left( C^\gamma_L \right)_{\beta \alpha} 
 & =
 \frac{2 s^2_W m^2_W}{16 \pi^2 m^2_{S_A}} \left( \xi^\dagger_1 {\bf I_\gamma} \xi_1\right)_{\beta \alpha} \,, \\
 \left( C^\gamma_R \right)_{\beta \alpha} 
 & =
 \frac{2 s^2_W m^2_W}{16 \pi^2 m^2_{S_A}} \left( \xi^\dagger_4 {\bf I_\gamma} \xi_4\right)_{\beta \alpha} \,.
 \end{split}
 \end{equation}
The loop integrals for up-type quarks can be expressed by a $3\times 3$ diagonal matrix, denoted as ${\bf I}_\gamma={\rm diag}(I_\gamma(r^u_A), I_\gamma(r^c_A), I_\gamma(r^t_A))$ with $r^f_A=m^2_f/m^2_{S_A}$ and 
  \begin{equation}
I_\gamma(r) = \frac{2-r (7-11 r)}{108 (1-r)^3} + \frac{r^3\ln r }{18 (1-r)^4}\,.
  \end{equation}
The effective Hamiltonian from the $Z$-penguin diagrams is given by
 \begin{equation}
{\cal H}^Z_{\ell_\alpha \to 3 \ell_\beta} 
= 
2\sqrt{2} G_F 
\left[ \bar u_\beta \left( C^Z_L P_L + C^R_Z P_R\right) u_\alpha \right]
\left[ \bar u_\beta (c^\ell_V - c^\ell_A \gamma_5) v_\beta \right]
\,,
 \end{equation}
where the $Z$ couplings to leptons in the SM are $c^\ell_A=-1/2$ and $c^\ell_V=-1/2 + 2 s^2_W$. The coefficients $C^Z_{L,R}$ can be expressed as:
 \begin{equation}
 \begin{split}
C^Z_L & = \frac{1}{32 \pi^2} \frac{m^2_t}{m^2_{S_A}} I_Z(r^t_{S_A})(\xi^\dagger_1)_{\beta 3} (\xi_1)_{3\alpha} \,, \\
C^Z_R & = -\frac{1}{32 \pi^2}  \frac{m^2_t}{m^2_{S_A}} I_Z(r^t_{S_A})(\xi^\dagger_4)_{\beta 3} (\xi_4)_{3\alpha} \,,
  \end{split}
 \end{equation}
and the loop integral $I_Z$ is defined as:
\begin{equation}
I_Z(r) = - \frac{1}{1-r} - \frac{\ln r}{(1-r)^2}\,.
\end{equation}

Since the Yukawa couplings of charged leptons to up-type quarks and LQ $S_1$ involve the left- and right-handed states, the four-fermion effective operators contain vector- and scalar-type currents.  However, due to the chirality flip from the up-type quark mass, scalar-type operators are associated with a factor of $m_{u_k} m_{u_{k'}}/m^2_{S_A}$.  Even considering the dominant top-quark contribution, this factor is ${\cal O}(10^{-2})$.  Thus, compared to vector-type operators, the contributions of scalar-type operators are negligible.  The effective Hamiltonian for $\ell_\alpha \to 3 \ell_\beta$ decays from box diagrams is therefore written as 
 \begin{equation}
 \begin{split}
 {\cal H}_{\rm Box} 
 = & 
 2\sqrt{2} G_F \left[ \bar u_\beta \gamma_\mu \left( (C^{\rm box}_L)_{\beta \alpha} P_L + (C^{\rm box}_R)_{\beta\alpha} P_R \right) u_\alpha \right]
 \\
 & \times \left[ \bar u_\beta \gamma^\mu \left( (C^{\rm box}_L)_{\beta \beta} P_L + (C^{\rm box}_R)_{\beta \beta} P_R \right) v_\beta \right]
 \,,
 \end{split}
 \end{equation}
where
 \begin{equation}
\left( C^{\rm box}_L \right)_{\beta \alpha} 
= 
\frac{1}{2g^2 } \left(\frac{m^2_W }{16\pi^2  m^2_{S_A} } \right) \left( \xi^\dagger_1 \xi_1 \right)_{\beta \alpha}
\,,~~ 
\left( C^{\rm box}_R \right)_{\beta \alpha}
= 
\frac{1}{2 g^2}  \left( \frac{m^2_W }{16\pi^2  m^2_{S_A} } \right)
\left( \xi^\dagger_4 \xi_4 \right)_{\beta \alpha}\,. 
\label{eq:CLR_box}
 \end{equation}
Note that the effect of ${\cal O}(m^2_t/m^2_{S_A})$ has been neglected.  Therefore, the loop integral in Eq.~(\ref{eq:CLR_box}) reduces to a constant of $1/2$.

\subsection{$\mu-e$ conversion in nuclei}

The $\mu-e$ conversion process describes the capture of a muon by a nucleus through the process $\mu (A,Z) \to e (A,Z)$.  At the quark level, this conversion occurs via the process $\mu q \to e q$, mediated by tree-level diagrams involving $S_A$ in the model. Although photon- and $Z$-penguin diagrams also contribute to this process, their effects are significantly smaller than those of the tree-level diagrams. Therefore, we focus on the tree-level contributions. Using the couplings from Eq.~(\ref{eq:Leptoquarkyukawas}), the effective Hamiltonian for $\mu-e$ conversion is
\begin{equation}
\begin{split}
{\cal H}_{\mu-e} 
=& 
-\frac{1}{2 m^2_{S_A}} \Bigg[ 
(\xi_1)_{i2} (\xi^\dag_1)_{1j} \overline{u_L}_j \gamma_\mu u_{Li} \bar e_L \gamma^\mu \mu_L 
+ (\xi_4)_{i2} (\xi^\dag_4)_{1j} \overline{u_L}_j \gamma_\mu u_{Li} \bar e_L \gamma^\mu \mu_L 
\\
&+ (\xi_1)_{i2} (\xi^\dag_4)_{1j} \left(  -\overline{u_R}_j  u_{Li} \bar e_R  \mu_L + \frac{1}{4} \overline{u_R}_j  \sigma_{\mu \nu}u_{Li} \bar e_R  \sigma^{\mu\nu} \mu_L \right) 
\\
& + (\xi_4)_{i2} (\xi^\dag_1)_{1j} \left(  -\overline{u_L}_j  u_{Ri} \bar e_L  \mu_R + \frac{1}{4} \overline{u_L}_j  \sigma_{\mu \nu}u_{Ri} \bar e_L  \sigma^{\mu\nu} \mu_R \right)
\Bigg] ~.
\end{split}
\end{equation}
In this model, the charged leptons couple only to the up-type quarks, here specifically the $u$- and $c$-quarks present in the nuclei. As in other LFV processes, the relevant Yukawa couplings are $\xi_1$ and $\xi_4$.  Furthermore, Fierz transformations introduce not only vector currents but also scalar and tensor currents into the effective Hamiltonian.

\subsection{$\Delta F=2$ and the $b\to u_j \ell \bar\nu_{\ell'}$ decays}

The $\Delta F=2$ process arises from box diagrams.  In addition to the two LQs, additional mediators can be neutrinos for the $B_q$- and $K$-meson systems and charged leptons for the $D$-meson system.  Since the couplings involving charged leptons are stringently constrained by LFV processes, we focus on the $\Delta B=2$ and $\Delta K=2$ processes.  In such cases, we have the $\Delta B=2$ and $\Delta S=2$ effective Hamiltonian
\begin{align}
{\cal H}^{NP} _{\Delta F=2}
= & 
\frac{1}{128\pi^2 m^2_{S_A}} 
\left[\bar d_j \gamma_\mu (\tilde{\xi}_2 \tilde{\xi}^\dagger_2)^*_{ji} P_L d_i \right] 
\left[ \bar d_j \gamma^\mu (\tilde{\xi}_2 \tilde{\xi}^\dagger_2)^*_{ji} P_L d_i\right]  
\nonumber \\ 
& + \frac{1}{128\pi^2 m^2_{S_B}} 
\left[\bar d_j \gamma_\mu (\tilde{\xi}_3 \tilde{\xi}^\dagger_3)^*_{ji} P_R d_i \right] 
\left[ \bar d_j \gamma^\mu ( \tilde{\xi}_3  \tilde{\xi}^\dagger_3)^*_{ji} P_R d_i\right]
~,
\label{eq:HdF=2}
\end{align}
where $j=3$ and $i=1(2)$ for the $B_{d(s)}$ meson and $j=2$ and $i=1$ for the $K$ meson.  Accordingly, the $F-\bar F$ mixing matrix element is given by
 \begin{equation}
 M^{F}_{12} 
 = 
 \langle \bar F| {\cal H}^{\rm SM}_{\Delta F=2} 
 + {\cal H}^{NP}_{\Delta F=2}  | F \rangle 
 = 
 |M^{F}_{12} | e^{i\phi_F}\,.
 \end{equation}
The mixing parameters in the $B$ and $K$ mesons can then be estimated via
 \begin{equation}
 \begin{split}
 \Delta m_{B_q} & \approx 2 | M^{B_q}_{12}|\,, 
 \\
 \Delta m_K & \approx  2 {\rm Re}(M^{K}_{12})\,.
 \end{split}
 \end{equation}
Furthermore, using the time-dependent CP asymmetry and choosing appropriate decay modes, the CP asymmetries induced by $F-\bar F$ oscillations for $B_q$ and $K$ mesons can be respectively expressed as~\cite{ParticleDataGroup:2024cfk}:
  \begin{equation}
  \begin{split}
  S^{B_q}_f &\simeq - \sin \phi_{B_q}
  \,, \\
  |\epsilon_K| & \simeq \frac{1}{\sqrt{2}} \frac{Im M^K_{12} }{\Delta m_K}~,
  \end{split}
  \end{equation}
where the decay mode $f$ could be $J/\Psi K_S$ for $B_d$ and $J/\Psi \phi$ for $B_s$.

Using the Yukawa couplings in Eq.~(\ref{eq:Leptoquarkyukawas}), the effective Hamiltonian for the semileptonic $b\to u_j \ell \bar\nu_{\ell'}$ decays is derived as:
  \begin{equation}
   \begin{split}
   {\cal H}_{b\to u_j \ell \bar\nu} 
   =  
   \frac{1}{2 m^2_{S_A}} & 
   \Big[ - (\tilde{\xi}^T_2)_{\ell' 3}  (\xi^*_1)_{ j \ell}\, \bar u_{L j} \gamma^\mu b_{L} \,\bar\ell_L \gamma_\mu \nu_{\ell' L} 
   + (\tilde{\xi}^T_2)_{\ell' 3}  (\xi^*_4)_{ j \ell}\, \bar u_{R j}  b_{L} \,\bar\ell_R   \nu_{\ell' L} 
   \\
 &~ - \frac{1}{4}  (\tilde{\xi}^T_2)_{\ell' 3}  (\xi^*_4)_{ j \ell}\, \bar u_{R j}  \sigma_{\mu \nu}  b_{L} \,\bar\ell_R  \sigma^{\mu \nu} \nu_{\ell' L}  \Big]~.
\label{eq:H_buell}
   \end{split}
  \end{equation}
It is known that the right-handed neutrinos couple only to the down-type quarks and the LQ $S'$.  Therefore, the induced $b\to u_j \ell \bar\nu_{\ell'}$ decays are suppressed by the small mixing angle $\theta_S$.  As a result, Eq.~(\ref{eq:H_buell}) involves only the left-handed neutrinos, with the relevant couplings given by $(\tilde{\xi}^T_2)_{\ell' 3} (\xi^*_{1,4})_{j \ell}$, where $j=1$ for the $b\to u$ transition and $j=2$ for the $b\to c$ transition.  The scalar and tensor currents in the Hamiltonian arise from Fierz transformations.  To investigate the LQ effects on the $B\to D^{(*)} \tau \bar\nu$ decays, we consider the ratios
 \begin{equation}
 R_{D^{(*)}} 
 = 
 \frac{{\cal B} (B\to D^{(*)} \tau \bar\nu)}
      {{\cal B} (B\to D^{(*)} \ell \bar\nu)}\,,
 \end{equation}
with $\ell = e,\mu$.

\section{Numerical Analysis and Discussions} \label{sec:NA}

We present in this section a detailed numerical analysis, followed by phenomenological discussions.  To find the best values of the free parameters that fit experimental data, we adopt the minimum $\chi^2$ approach, where the weighted $\chi^2$ is defined as:
 \begin{equation}
 \chi^2 = \sum_i \frac{ (O^{\rm th}_i -O^{\rm exp}_i )^2} {\sigma^2_i}\,. \label{eq:chi2}
 \end{equation}
Here $O^{\rm th}_i $ and $O^{\rm exp}_i$ denote the model-predicted value and the measured central value for the $i$-th observable, respectively. The weight factor $\sigma^2_i = (\sigma^{\rm SM}_i)^2 + (\sigma^{\rm exp}_i)^2$ combines the uncertainties from both SM predictions and experimental data in quadrature.

\subsection{Parameter setting and numerical inputs} \label{subsec:para_inputs}

The flavor-dependent Yukawa couplings in Eq.~(\ref{eq:Leptoquarkyukawas}) involve many new free parameters, which complicate the analysis.  To reduce the number of free parameters, we adopt the orthogonal matrix instead of the unitary matrix $U$ as in Eq.~(\ref{eq:para_1}) for $\tilde{\xi}_{2,3}$.  The parametrization with three Euler angles is given by
   \begin{equation}
   U=
\left(
\begin{array}{ccc}
 \cos\theta_{12} & -\sin\theta_{12} & 0 \\
 \sin\theta_{12} & \cos\theta_{12} & 0 \\
 0 & 0 & 1 \\
\end{array}
\right).\left(
\begin{array}{ccc}
 \cos\theta_ {13} & 0 & -\sin\theta_{13} \\
 0 & 1 & 0 \\
 \sin\theta_{13} & 0 & \cos\theta_{13} \\
\end{array}
\right).\left(
\begin{array}{ccc}
 \cos\theta_{23} & \sin\theta_{23} & 0 \\
 -\sin\theta_{23} & \cos\theta_{23} & 0 \\
 0 & 0 & 1 \\
\end{array}
\right)~,
   \end{equation}
where the ranges of the angles are taken as $\theta_{12} \in [0, \pi)$, $\theta_{13} \in [-\pi/2, \pi/2)$, and $\theta_{23} \in (-\pi, \pi]$~\cite{2018EPJP..133..206G}.  Additionally, the matrix $\epsilon$ in Eq.~(\ref{eq:para_1}) is assumed to be real.  Since neutrino oscillation experiments cannot determine the absolute neutrino masses, we assume the lightest neutrino mass to be zero: $m_{\nu 1}=0$ for the normal mass ordering and $m_{\nu 3}=0$ for the inverted mass ordering.  We have verified that this assumption is consistent with the results of a $\chi^2$ analysis when the lightest mass is treated as a free parameter.  As a result, $\varepsilon_{1(3)}$ in Eq.~(\ref{eq:e_para}) becomes an irrelevant parameter when $m_{\nu_{1(3)}}=0$.  To achieve a neutrino mass scale on the order of $10^{-3}-10^{-2}$~eV, the LQ mixing angle is fixed at $\theta_{S} = 8\times 10^{-8}$.  In addition to $\xi_1$, the LFV processes and $b\to u_j \ell \nu$ involve the $\xi_4$ matrix, which is completely unknown.  To simplify the numerical analysis, we set $(\xi_4)_{1\ell} = (\xi_4)_{u_j 1} \approx 0$.  To satisfy the constraints from LFV processes, the ranges for the remaining parameters are taken as:
\begin{equation}
\begin{split}
 \zeta_{11} & \equiv (\xi_{4})_{c\mu} \in (-10^{-4},\, 10^{-4})~, \\
 \zeta_{21} & \equiv (\xi_{4})_{t \mu} \in (-10^{-4}, \, 10^{-4})~,  \\
 \zeta_{22} & \equiv (\xi_4)_{t\tau} \in (-0.5, \, 0.5)~.
 \end{split}
 \end{equation}
Using the high-$p_T$ tail of the $pp\to \tau\tau$ distribution measured by ATLAS~\cite{ATLAS:2020zms}, a bound on the $c$-$\tau$-$S_1$ coupling was derived as $|(\xi_4)_{t\tau}|\lesssim 1.6$~\cite{Angelescu:2021lln}. Based on this result, we use $\zeta_{12}=(\xi_4)_{t\tau}=1.5$ in the numerical estimates, fixing the LQ masses $m_{S_A}=1.5$~TeV and $m_{S_B}=3$~TeV.  In summary, there are eight free parameters, including three Euler angles, two $\varepsilon_i$, and three $\zeta_{11,21,22}$, involved in $\xi_1$ and $\tilde{\xi}_{2,3}$.

To apply the parametrization of $\tilde{\xi}_{2,3}$ defined in Eq.~(\ref{eq:para_1}) and $\xi_1$ defined in Eq.~(\ref{eq:xi_1}), it is necessary to implement the values of the PMNS matrix elements and the neutrino masses.  For this purpose, we refer to the latest global fitting results with Super-Kamiokande (SK) atmospheric data as presented in Ref.~\cite{Esteban:2024eli}.  The values with $1\sigma$ errors for normal mass ordering are as follows:
\begin{equation}
\begin{split}
\phi_{12}/^{\circ} & = 33.68^{+0.73}_{-0.70},~~ 
\phi_{23}/^{\circ}=43.3^{+1.0}_{-0.8},~~ 
\phi_{13}/^{\circ}=8.56^{+0.11}_{-0.11},~~ 
\delta_{CP}/^{\circ}=212^{+26}_{-41}   , 
\\
\Delta m^2_{21}/\text{eV$^2$}& = (7.49^{+0.19}_{-0.19}) \times 10^{-5},~~ 
\Delta m^2_{3\ell}/\text{eV$^2$}= (2.513^{+0.021}_{-0.019})\times 10^{-3},
\end{split}
\end{equation}
where $\phi_{12,13,23}$ are the mixing angles in the PMNS matrix, $\delta_{CP}$ is the Dirac CP-violating phase, and $\Delta m^2_{3\ell}$ represents the mass-square difference relative to the lightest neutrino.  For the inverted mass ordering, the corresponding central values are:
 \begin{equation}
\begin{split}
\phi_{12} /^{\circ}& = 33.68^{+0.73}_{-0.70},~~ 
\phi_{23}/^{\circ}=47.9^{+0.7}_{-0.9},~~ 
\phi_{13}/^{\circ}=8.59^{+0.11}_{-0.11},~~ 
\delta_{CP}/^{\circ}=274^{+22}_{-25}   , 
\\
\Delta m^2_{21}/\text{eV$^2$} & = (7.49^{+0.19}_{-0.19}) \times 10^{-5},~~ 
\Delta m^2_{3\ell}/\text{eV$^2$}= (-2.484^{+0.020}_{-0.020})\times 10^{-3} ~.
\end{split}
\end{equation}

To constrain the parameters $\zeta_{11,21,22}$, which would influence the LFV processes, we rely on the relevant experimental upper limits shown in Table~\ref{tab:LFV}.  As discussed earlier, the numerical analysis involves eight selected free parameters.  To determine their optimal values, we use nine observables for the $\chi^2$ evaluation, leaving one degree of freedom.  The relevant processes, along with their SM predictions and the corresponding experimental data, are presented in Table~\ref{tab:inputs}.   Note that since the influence on $\Delta m_K$ is small, we do not include it in the analysis.  Because $\tilde{\xi}_{2,3}$ are real in this study, the LQs have no influence on $K_L\to \pi^0 \nu \bar\nu$ and we focus the discussion on the $K^+\to \pi^+ \nu \bar\nu$ process.

\begin{table}[thp]
\caption{Current experimental upper limits (EULs) on LFV processes.}
\begin{center}
\begin{tabular}{c|cccc}
\hline \hline
 Obs. ~& ~${\cal B}(\mu\to e \gamma)\cdot 10^{13}$ ~& ~ ${\cal B}(\mu\to 3e)\cdot 10^{12}$ ~  & ~ ${\cal B}(\tau\to e \gamma) \times 10^{8}$~&~${\cal B}(\tau\to \mu \gamma) \times 10^{8}$ 
 \\ \hline
 EUL ~&~ $<4.2$~\cite{MEG:2016leq} 
 & $< 1.0$~\cite{SINDRUM:1987nra}  & $< 3.3 $~\cite{BaBar:2009hkt}  
 & $< 4.2$~\cite{Belle:2021ysv} 
 \\  \hline \hline 
 Obs. ~& ~${\cal B}(\tau \to 3e)\cdot 10^{8}$ ~& ~ ${\cal B}(\tau\to 3\mu)\cdot 10^{8}$ ~  & ~ $CR(\mu-e, \rm{Au}) \times 10^{13}$~&~$CR(\mu-e, \rm{Ti}) \times 10^{12}$ 
 \\ \hline
 EUL ~&~ $< 2.7$~\cite{Hayasaka:2010np} 
 & $< 2.1$~\cite{Hayasaka:2010np}&  $< 7$~\cite{SINDRUMII:2006dvw} 
 & $< 4.3$~\cite{SINDRUMII:1993gxf}
 \\ \hline\hline
\end{tabular}
\end{center}
\label{tab:LFV}
\end{table}

\begin{table}[thp]
\caption{The experimental measurements and the SM predictions used in the $\chi^2$ fit.}
\begin{center}
\begin{tabular}{c|ccc}
\hline \hline
 Obs. ~& ~$\Delta m_{B_d} \cdot 10^{13}$ [GeV]~& ~ $\Delta m_{B_s} \cdot 10^{12}$ [GeV]~  & ~ $S_{J/\Psi K^0}$ 
 \\ \hline
 Exp.~\cite{ParticleDataGroup:2024cfk}~ & $3.3349\pm 0.0125$ & $11.688\pm 0.003$  & $0.708 \pm 0.017 $  
 \\ 
 SM~\cite{Straub:2018kue}~& $3.496\pm 0.258$  & $11.560\pm 0.063$  & $0.708\pm 0.026$  
 \\  \hline \hline 
 Obs.  ~&~$S_{J/\Psi \phi}$~& ~ ${\cal B}(K^+ \to \pi^+ \nu \bar\nu)\cdot 10^{10}$ ~& ~${\cal B}(B^+ \to K^+ \nu \bar\nu) \cdot 10^{5}$
 \\ \hline
 Exp. ~&~ $0.036\pm 0.016$~\cite{ParticleDataGroup:2024cfk}  &  $1.14^{+0.40}_{-0.33}$~\cite{ParticleDataGroup:2024cfk}  & $1.3 \pm 0.4$~\cite{Belle-II:2023esi}  
 \\ 
 SM  ~&~ $0.0369\pm 0.0017$~\cite{Straub:2018kue}  & $0.86\pm 0.042$~\cite{Buras:2022qip} & $0.492 \pm 0.03$~\cite{Buras:2022qip}  
 \\ \hline\hline
 Obs.  ~& ~ ${\cal B}(B\to \tau \nu)\cdot 10^{4}$~&~ $R_D$~ & ~$R_{D^*}$
 \\ \hline
 Exp. ~&~ $1.09\pm 0.24$~\cite{ParticleDataGroup:2024cfk}  &  $0.342\pm 0.026$~\cite{HeavyFlavorAveragingGroupHFLAV:2024ctg}  &  $0.287\pm 0.012~$~\cite{HeavyFlavorAveragingGroupHFLAV:2024ctg} 
 \\ 
 SM~\cite{Straub:2018kue} ~&~ $0.84\pm 0.07$
 &  $0.2950 \pm 0.0055$ & $0.2451\pm 0.0066$  \\ \hline \hline
\end{tabular}
\end{center}
\label{tab:inputs}
\end{table}

\subsection{Optimized results from $\chi^2$ minimization}

Based on the parameter setup and the adopted values in the previous subsection, we present the numerical results obtained by minimizing $\chi^2$ defined in Eq.~(\ref{eq:chi2}).  In Eq.~(\ref{eq:e_para}), we introduce six simple patterns for each of the $\epsilon$ and $\eta$ matrices within our parametrization of $\tilde{\xi}_{2}$ and $\tilde{\xi}_{3}$.  To identify the pattern that enhances $R^{\nu\nu}_K$, $R^{\nu\nu}_{\pi^+}$, and $R_D$, we compute all possible combinations of $\epsilon_i$ and $\eta_j$.  The numerical results for each $(\epsilon_i, \eta_j)$ combination under both normal and inverted ordering scenarios are presented in Tables~\ref{tab:NO} and \ref{tab:IO}, respectively. In addition to the minimum $\chi^2$ values, these tables also present the predicted values of various observables: ${\cal B}(K^+\to \pi^+ \nu \bar\nu)$, ${\cal B}(B\to K \nu \bar\nu)$, ${\cal B}(\mu\to e \gamma)$, ${\cal B}(\tau \to \ell \gamma)$, and the capture rate for $\mu$-$e$ conversion in a gold nucleus.  Since not all combinations of $(\epsilon_i, \eta_j)$ satisfy the current experimental upper limits on LFV processes, we only present those patterns that are allowed by experimental data.  Across all the patterns, $R_{D^*}$ is less affected and thus not shown in the tables.  It is evident in both tables that while $B\to K \nu \bar\nu$ and $K^+\to \pi^+ \nu\nu$ are enhanced, the resulting $R_D$ remains unchanged in the inverted ordering scenario, where $\chi^2_{\rm min}$ is generally larger. Consequently, a more detailed analysis will be focused on the normal ordering scenario.

According to the results in Table~\ref{tab:NO}, we find that the pattern $(\epsilon_d, \eta_d)$ yields not only the smallest minimum $\chi^2$ value, $\chi^2_{\rm min}=8.03$, but also the largest $R_D$ of $0.323$. As a comparison, the SM has $\chi^2_{\rm min}=$18.3, with the main contributions coming from $R_{D^{(*)}}$ and $B \to K\nu \bar\nu$.  Although the pattern $(\epsilon_e, \eta_f)$ enhances $R_D$ up to the $1\sigma$ lower bound of the data, its $\chi^2_{\rm min}$ remains larger than that of $(\epsilon_d, \eta_d)$.  Therefore, we select the pattern $(\epsilon_d, \eta_d)$ as the representative case to illustrate the characteristics of the model.

\begin{table}[thp]
\caption{Predictions of observables based on the optimized parameters obtained from a $\chi^2$ fit for different $(\epsilon_i, \eta_j)$ combinations in the normal mass ordering scenario.  }
\begin{center}
\begin{tabular}{c|cccccc}
\hline \hline
 $(\epsilon, \eta)$ ~& ~$(\epsilon_a,\, \eta_e)$ ~& ~$(\epsilon_b,\, \eta_e)$  ~&  ~$(\epsilon_c,\, \eta_c)$~ &~$(\epsilon_d, \eta_d)$~ & ~$(\epsilon_e,\, \eta_f)$~ & ~$(\epsilon_f,\, \eta_e)$ 
 \\ 
 $\chi^2_{\rm min}$ & $10.7$ & $12.9$ &  $10.8$ &  $8.03$ & $9.64$ & $16.3$
 \\ \hline
 $R_D$ & $0.309$  & $0.307$ & $0.310$ &  $0.323$ & $0.316$ & $0.308$
 \\ 
${\cal B}(K^+\to \pi^+ \nu \bar\nu)\cdot 10^{-10}~$ & $1.05$  & $1.00$ & $1.04$ &  $1.07$ & $1.31$ & $0.92$
 \\ 
 ${\cal B}(B^+\to K^+ \nu \bar\nu)\cdot 10^{-5}$ ~&  $1.09$  & $1.29$ & $1.13$ &  $1.06$ & $0.97$ & $1.8$
 \\ 
 ${\cal B}(\mu\to e\gamma)\cdot 10^{13}$ &  $1.62$  & $1.97$ & $1.34$ &  $2.59$ & $0.66$ & $0.99$
 \\ 
 ${\cal B}(\tau \to e \gamma)\cdot 10^{8}$ &  $1.18$ &  $1.52$ & $0.93$ &  $0.80$ & $1.25$ & $0.77$
 \\ 
 ${\cal B}(\tau\to \mu \gamma)\cdot 10^{8}$ & $2.78$ & $0.16$ & $2.49$ &  $1.40$ & $0.09$ & $0.25$
 \\ 
 $CR(\mu-e, \rm{Au})\cdot 10^{13}$  & $0.05$ & $5.24$ & $6.07$ &  $1.04$ &  $2.19$ & $1.42$
\\ \hline\hline
\end{tabular}
\end{center}
\label{tab:NO}
\end{table}

\begin{table}[thp]
\caption{Predictions of observables based on the optimized parameters obtained from a $\chi^2$ fit for different $(\epsilon_i, \eta_j)$ combinations in the inverted mass ordering scenario.  }
\begin{center}
\begin{tabular}{c|cccccc}
\hline \hline
 $(\epsilon, \eta)$ ~& ~$(\epsilon_a,\, \eta_e)$ ~& ~$(\epsilon_b,\, \eta_e)$  ~&  ~$(\epsilon_c,\, \eta_d)$~ &~$(\epsilon_d, \eta_c)$~ &~$(\epsilon_e, \eta_e )$~& ~$(\epsilon_f,\, \eta_d)$ 
 \\ 
 $\chi^2_{\rm min}$ & $13.8$ & $13.6$ &  $14.0$ & $14.0$ & $13.9$ & $14.1$
 \\ \hline
 $R_D$ &  $0.296$ & $0.296$ & $0.295$  & $0.295$ & $0.295$ & $0.296$
 \\ 
${\cal B}(K^+\to \pi^+ \nu \bar\nu)\cdot 10^{-10}$~ & $1.20$ & $1.06$ & $1.19$ & $1.14$ & $1.34$ & $0.89$
 \\ 
 ${\cal B}(B^+\to K^+ \nu \bar\nu)\cdot 10^{-5}$ ~&  $1.24$  & $1.31$ & $1.32$ & $1.28$ & $1.24$ &  $1.43$
 \\ 
 ${\cal B}(\mu\to e\gamma)\cdot 10^{13}$ & $3.91$ & $3.65$ & $3.33$ & $0.68$ & $2.99$ & $2.93$
 \\ 
 ${\cal B}(\tau \to e \gamma)\cdot 10^{8}$ & $0.40$ & $1.52$ & $2.46$ & $0.29$ & $0.810$ & $1.20$
 \\ 
 ${\cal B}(\tau\to \mu \gamma)\cdot 10^{8}$ & $1.45$ & $0.07$ & $0.89$ & $1.73$ & $2.39$ & $0.12$
 \\ 
 $CR(\mu-e, \rm{Au})\cdot 10^{13}$ & $4.4$  & $5.34$ & $5.16$ & $3.14$ & $4.22$ & $1.94$
\\ \hline\hline
\end{tabular}
\end{center}
\label{tab:IO}
\end{table}

Using the pattern $(\epsilon_d, \eta_d)$ in the normal mass ordering scenario, the optimized parameters are summarized as:
 \begin{equation}
 \begin{split}
 \theta_{12}&=8.3\times 10^{-5},~~ 
 \theta_{13}=-0.04,~\theta_{23}=-3.14,~~ \varepsilon_2=0.116,~\varepsilon_3=1.19, 
 \\
 \zeta_{11}&=3.02\times 10^{-5},~~
 \zeta_{21}=1.04\times 10^{-5},~~
 \zeta_{22} = 1.98\times 10^{-2}~. 
 \label{eq:optimal_values}
 \end{split}
 \end{equation}
As discussed earlier, since the preferred lightest neutrino mass is set as zero, $m_{\nu 1}=0$, the parameter $\varepsilon_1$ becomes irrelevant.  We also show the predictions for the other observables based on the same model parameters as follows:
 \begin{equation}
 \begin{split}
 \Delta m_{B_s}&=11.50\times 10^{-12}~\text{GeV},~~
 \Delta m_{B_d}=3.46\times 10^{-13}~\text{GeV},~~
 S_{J/\Psi \phi}=0.036, 
 \\
S_{J/\Psi K^0}&=0.697,~~ 
{\cal B}(B\to \tau \nu)=8.41\times 10^{-5},~~
R_{D^*}=0.254.
 \end{split}
 \end{equation}
The predicted value of $R_{D^*}$ in this case is about $3\sigma$ lower than the experimental central value.  Since the effects contributing to $\mu\to 3 e$ and $\tau\to 3\mu$ (and similarly $\tau\to \mu e e$) are strongly correlated with those of the two-body radiative decays $\mu\to e \gamma$ and $\tau\to \mu \gamma$, their BRs in this model are several orders of magnitude smaller than the current upper limits.  Moreover, the predicted muon $g-2$ and EDM are of ${\cal O}(10^{-4})$ times smaller than the current experimental bounds.  Hence, we do not discuss them further.

\begin{figure}
    \centering
    \includegraphics[scale=0.4]{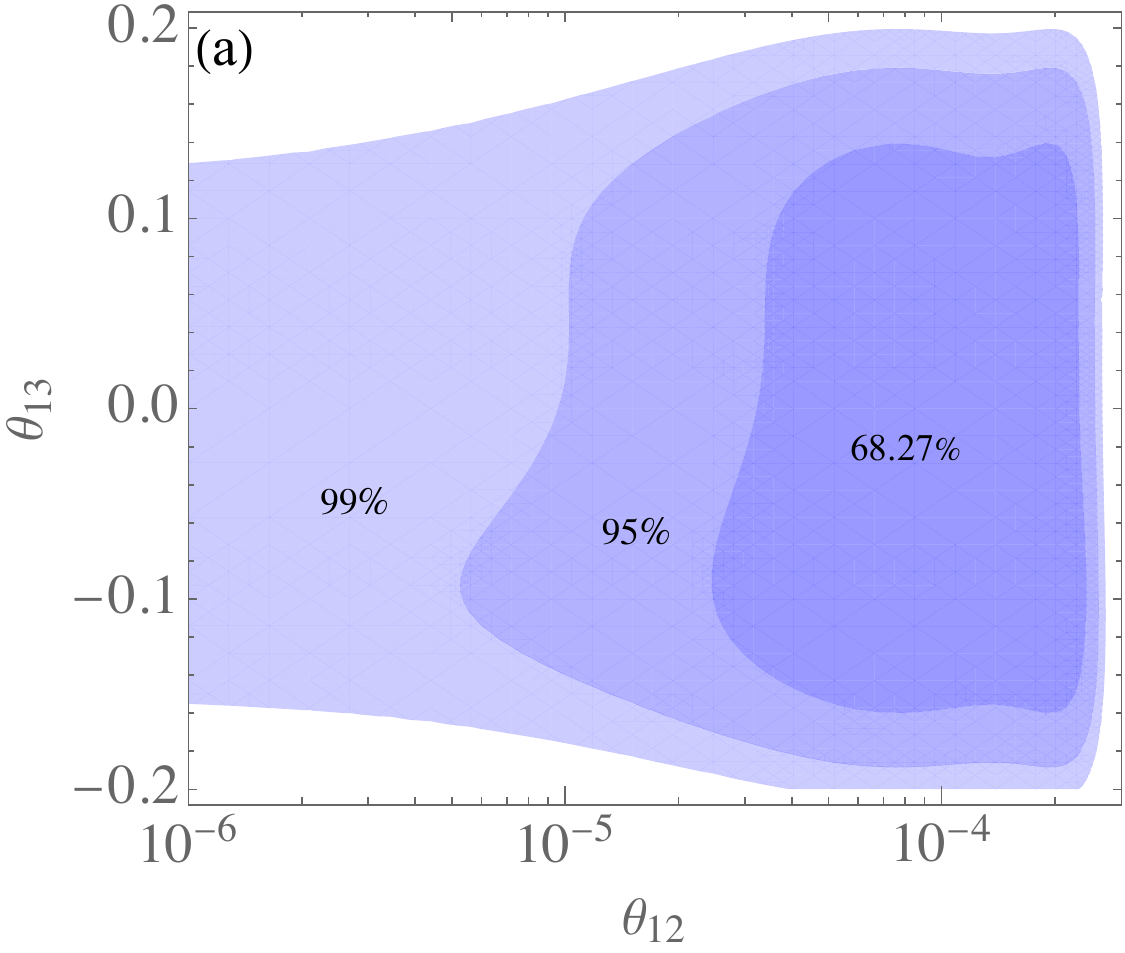}
    \hspace{2pt} 
    \includegraphics[scale=0.4]{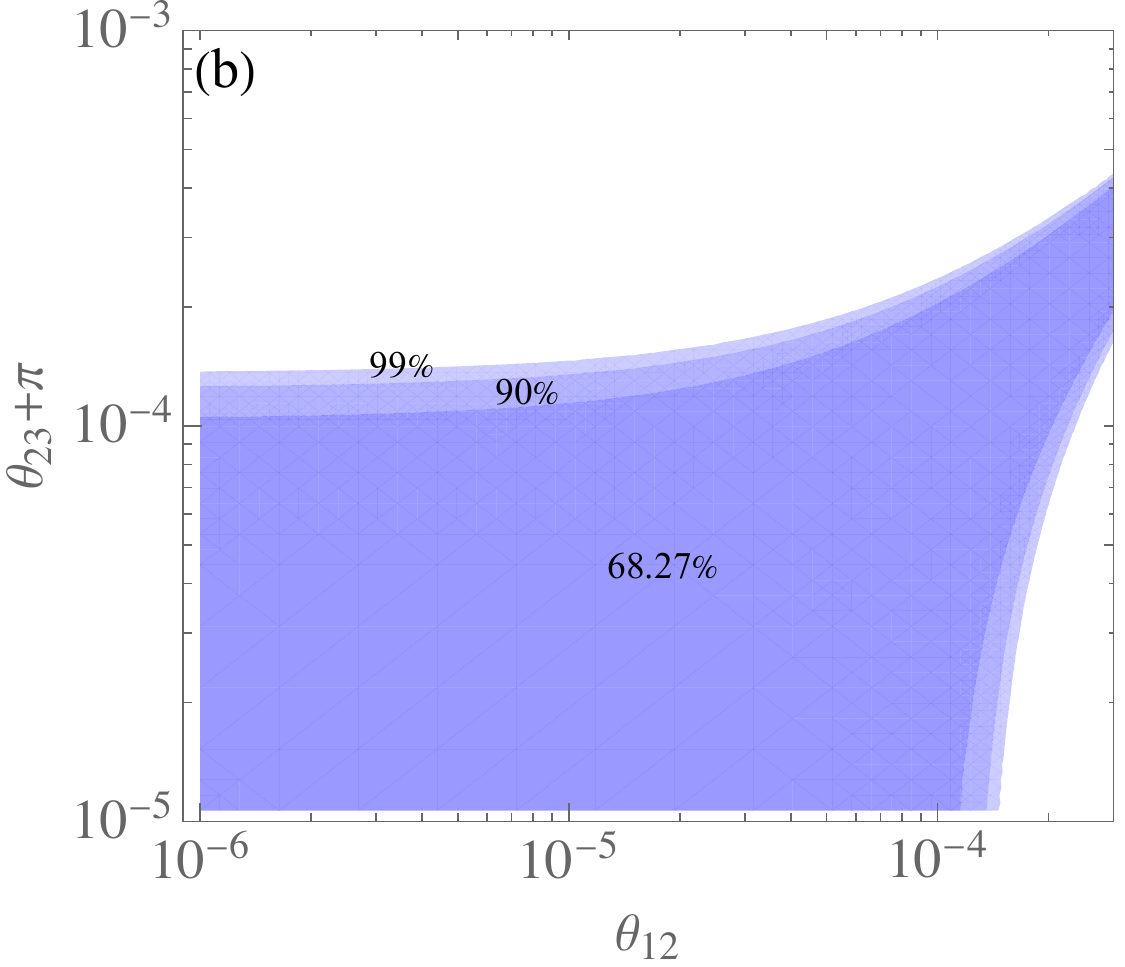} 
    \vspace{10pt}
    \\
    \includegraphics[scale=0.42]{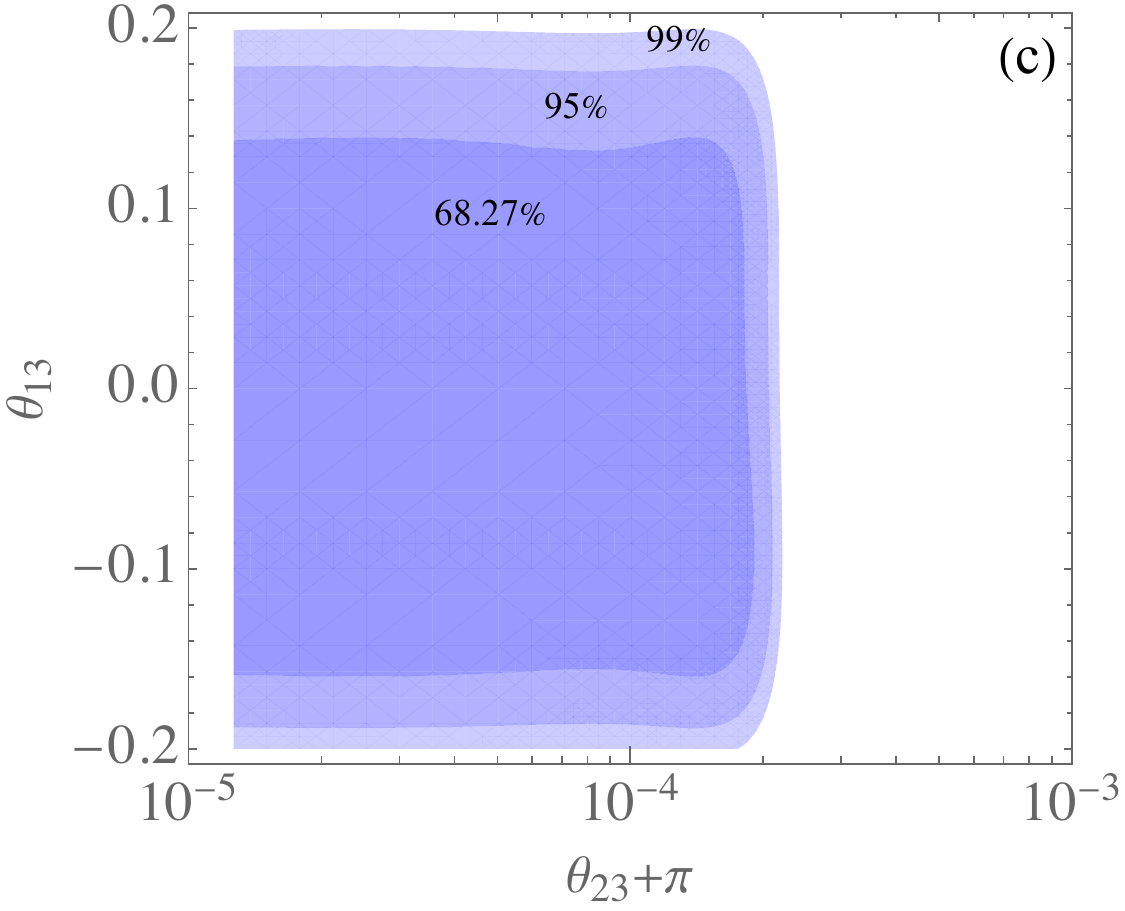}
    \hspace{2pt} 
    \includegraphics[scale=0.38]{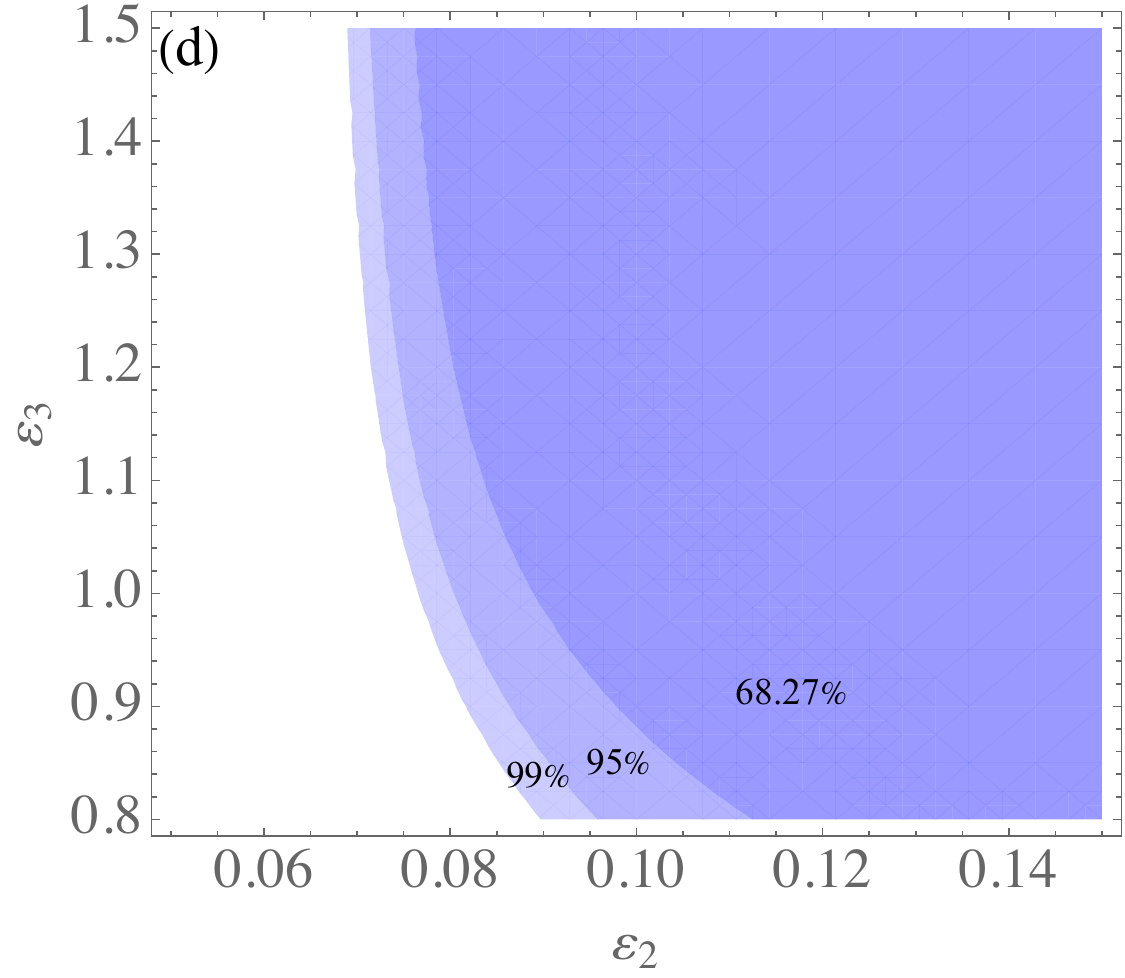}
    \caption{Contours of $\Delta\chi^2=\chi^2-\chi^2_{\rm min}$ in the plane of (a) $\theta_{12}$-$\theta_{13}$, (b) $\theta_{12}$-$\theta_{13}$, (c) $\theta_{23}$-$\theta_{13}$, and (d) $\varepsilon_2$-$\varepsilon_3$, with the $68.27\%$, $95\%$, and $99 \%$ CLs. }
    \label{fig:CL}
\end{figure}

To explicitly show the dependence on free parameters, we plot the $\Delta \chi^2=\chi^2-\chi^2_{\rm min}$ contours as functions of two selected parameters within the $68.27\%$, $95\%$, and $99 \%$ confidence levels (CLs) in Fig.~\ref{fig:CL}.  When varying these two parameters, all the other parameters are held fixed at their optimized values given in Eq.~(\ref{eq:optimal_values}).  From Fig.~\ref{fig:CL}, we observe that the preferred region for $\theta_{12}$ is constrained to small values, typically below ${\cal O}(10^{-4})$, while $\theta_{13}$ lies within $(-0.2, 0.2)$, and $\theta_{23} \simeq -\pi$.  Although $|\varepsilon_{2,3}|$ can reach up to 5 within the considered CLs, we choose to zoom in the ranges $\varepsilon_2\in (0.05, 0.15)$ and $\varepsilon_3 \in (0.8, 1.5)$, as this region contains the optimized parameters that yield $\chi^2_{\rm min}=8.03$ and satisfies the constraint from $\mu$-$e$ conversion at the same time.

\begin{figure}
    \centering
    \includegraphics[scale=0.4]{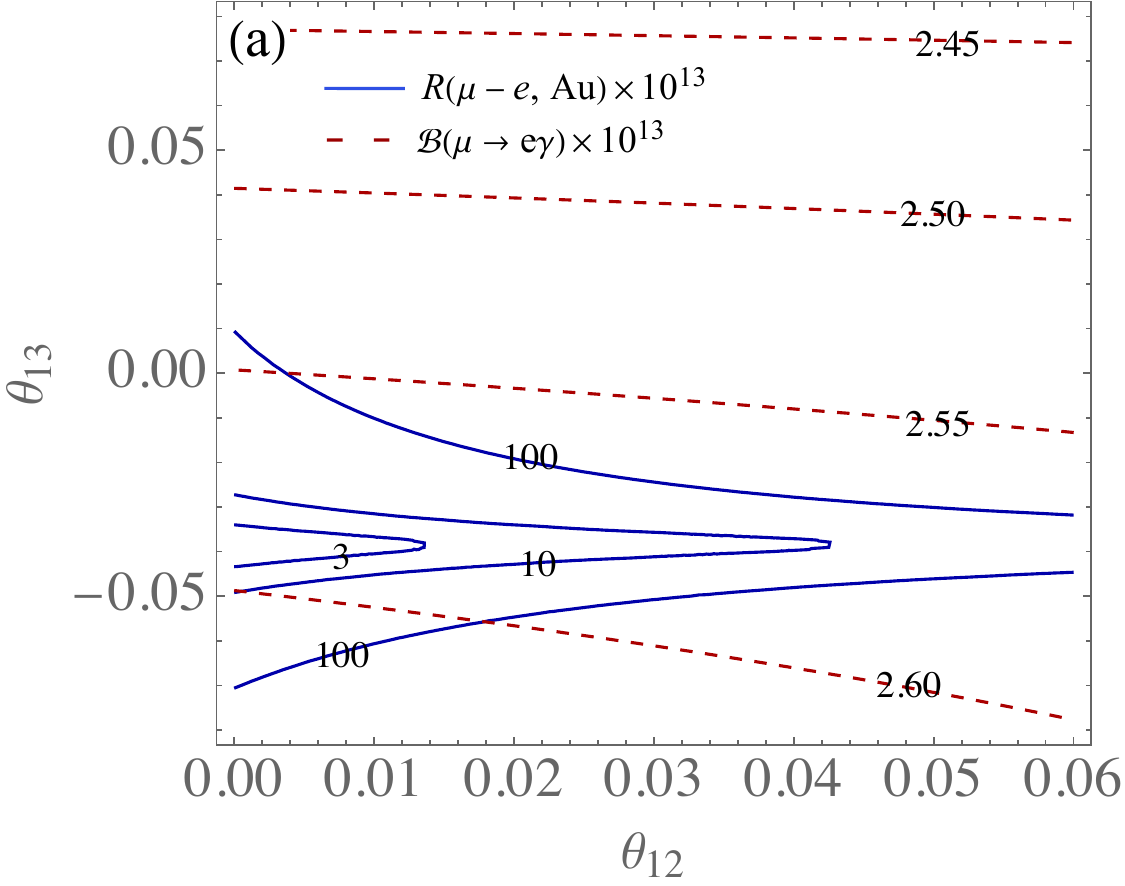}
    \hspace{0.5mm}
    \includegraphics[scale=0.4]{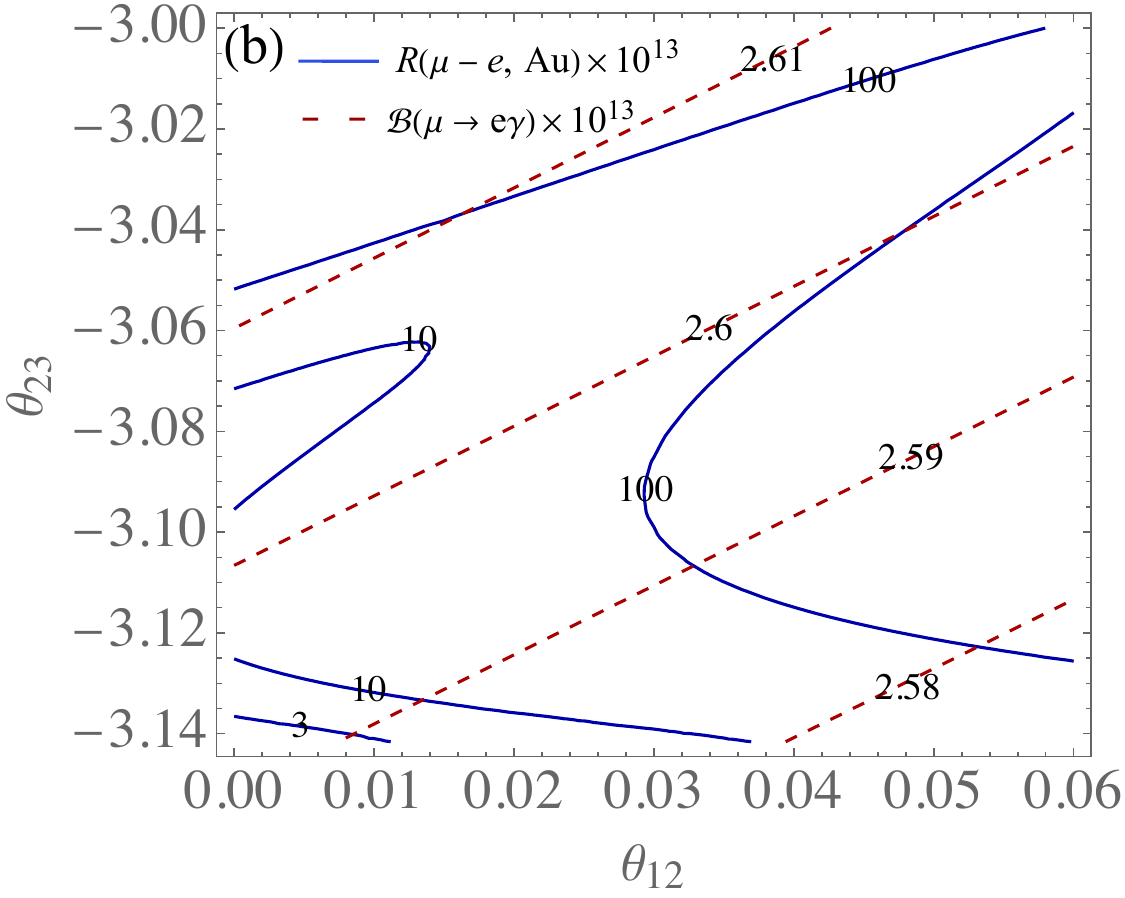} 
    \vspace{10pt}
    \\
    \includegraphics[scale=0.4]{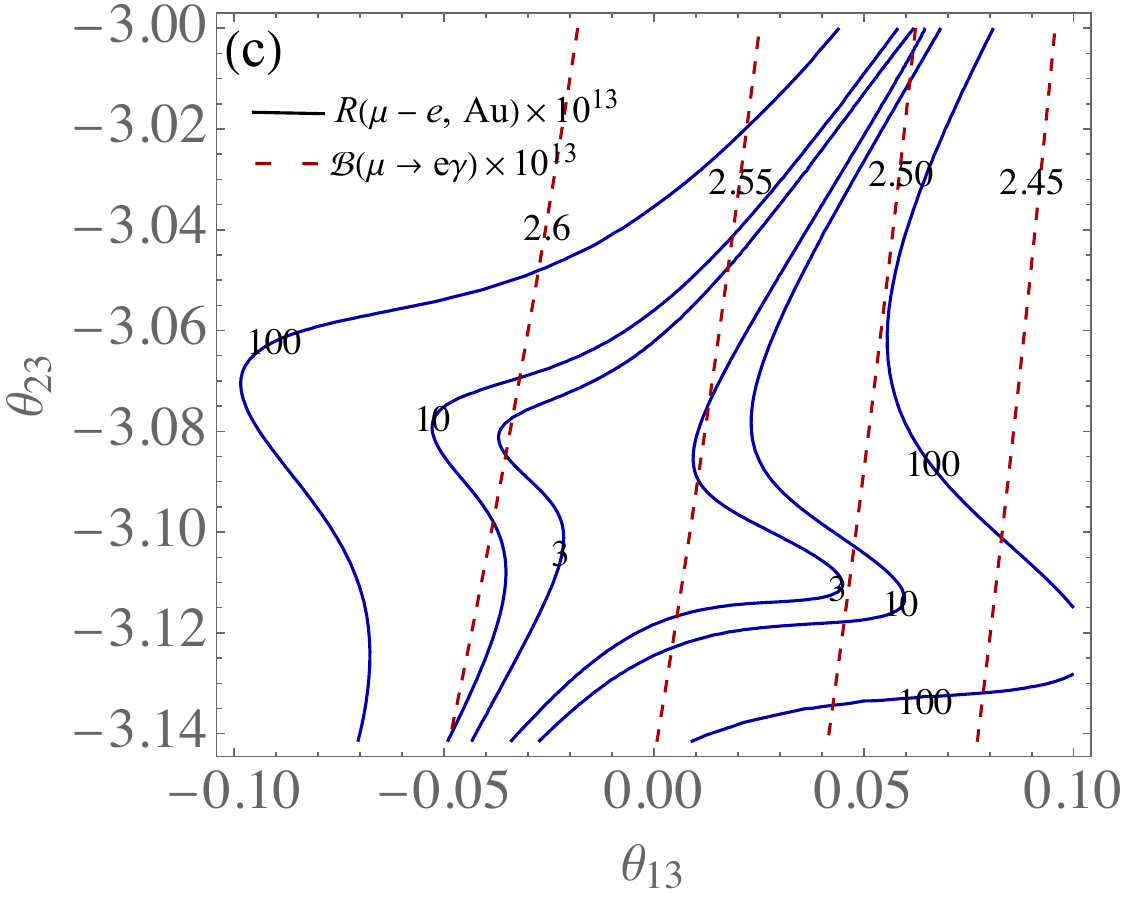}
    \hspace{2mm}
    \includegraphics[scale=0.38]{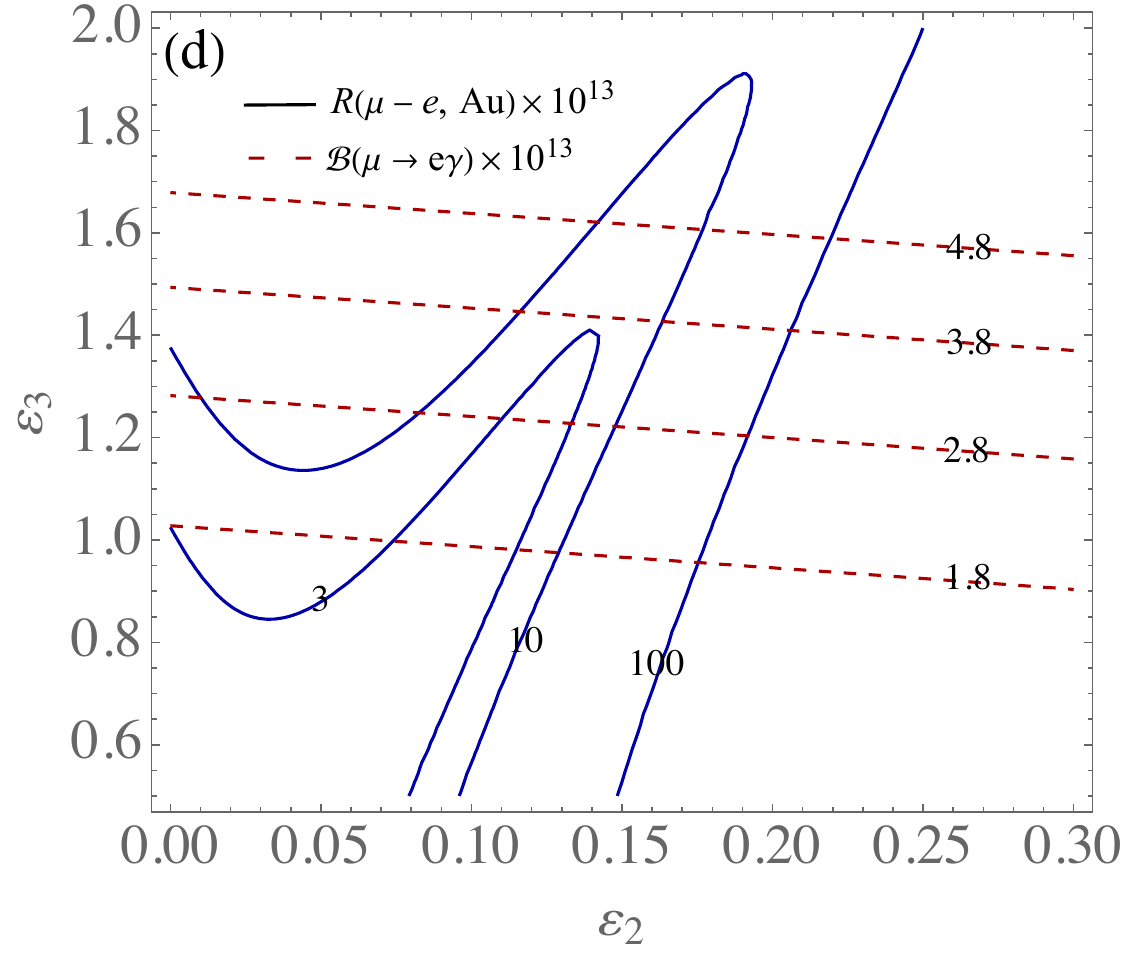}
    \caption{Contours of ${\cal B}(\mu\to e\gamma)$ (dashed) and $CR(\mu-e, \rm{Au})$ (solid) in the plane of (a) $\theta_{12}$-$\theta_{13}$, (b) $\theta_{12}$-$\theta_{13}$, (c) $\theta_{23}$-$\theta_{13}$, and (d) $\varepsilon_2$-$\varepsilon_3$. }
    \label{fig:mue_limit}
\end{figure}

The constraints on the parameters $\theta_{ij}$ and $\varepsilon_2$ primarily arise from the stringent constraints imposed by the tree-level induced $\mu-e$ conversion.  To illustrate this, we present the contour plots of $CR(\mu-e, \rm{Au})$ in the parameter planes of $\theta_{12}$-$\theta_{13}$, $\theta_{12}$-$\theta_{23}$, $\theta_{13}$-$\theta_{23}$, and $\varepsilon_{2}$-$\varepsilon_{3}$ in Fig.~\ref{fig:mue_limit}.  From Fig.~\ref{fig:mue_limit}(a), we see that the upper limit on $\mu$-$e$ conversion requires $\theta_{13} \sim -0.04$ and $\theta_{12}<0.01$. Similarly, Fig.~\ref{fig:mue_limit}(b) indicates that $\theta_{23} \simeq -\pi$, while Fig.~\ref{fig:mue_limit}(d) shows $\varepsilon_2 < 0.15$.  For comparison, we also include the BR of $\mu\to e\gamma$ in the same plots.  Due to loop-induced effects, the sensitivity of ${\cal B}(\mu\to e\gamma)$ on the parameters within the considered regions is weaker than that of the $\mu$-$e$ conversion.

\subsection{LQ contributions to $R^{\nu\nu}_K$, $R^{\nu\nu}_{\pi^+}$, and $R_D$}

\begin{figure}
    \centering
    \includegraphics[scale=0.4]{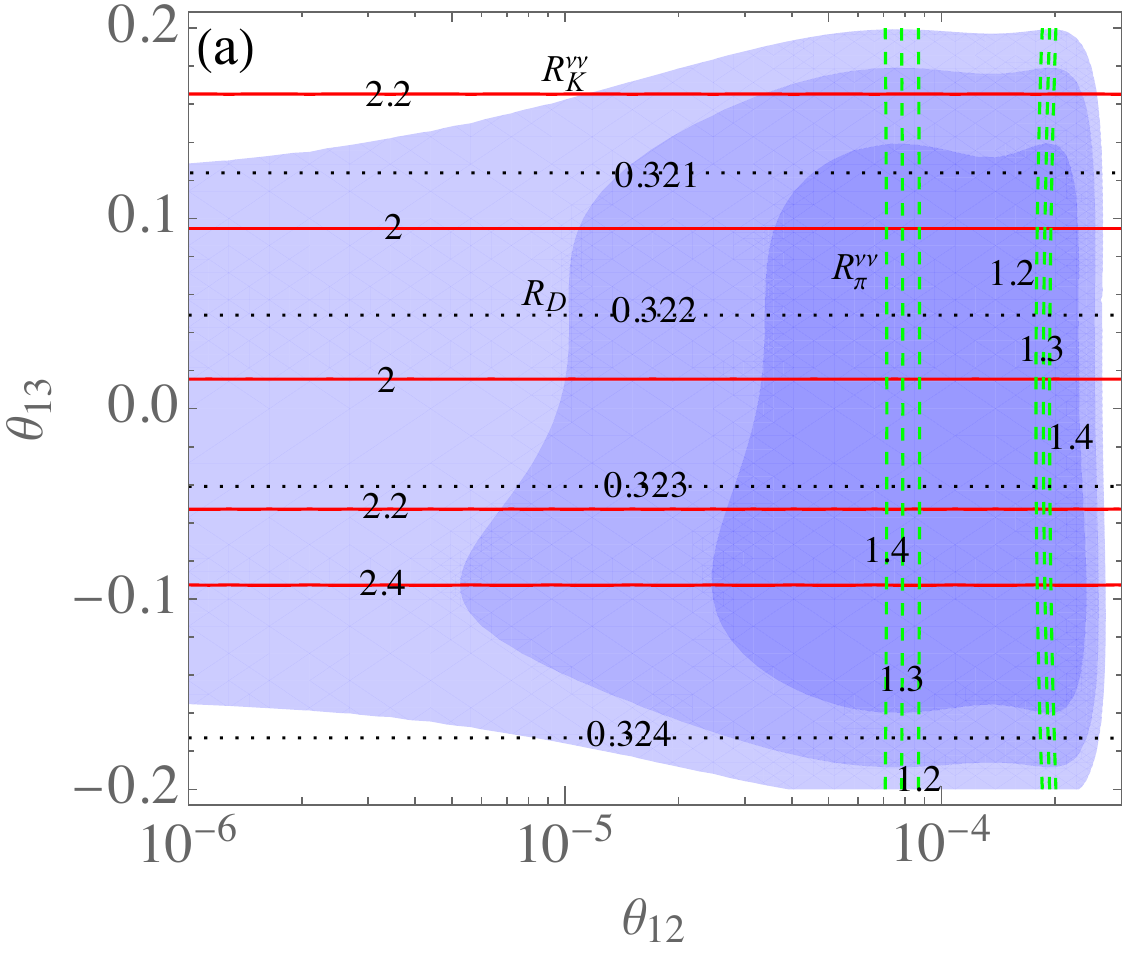}
    \hspace{1mm}
    \includegraphics[scale=0.4]{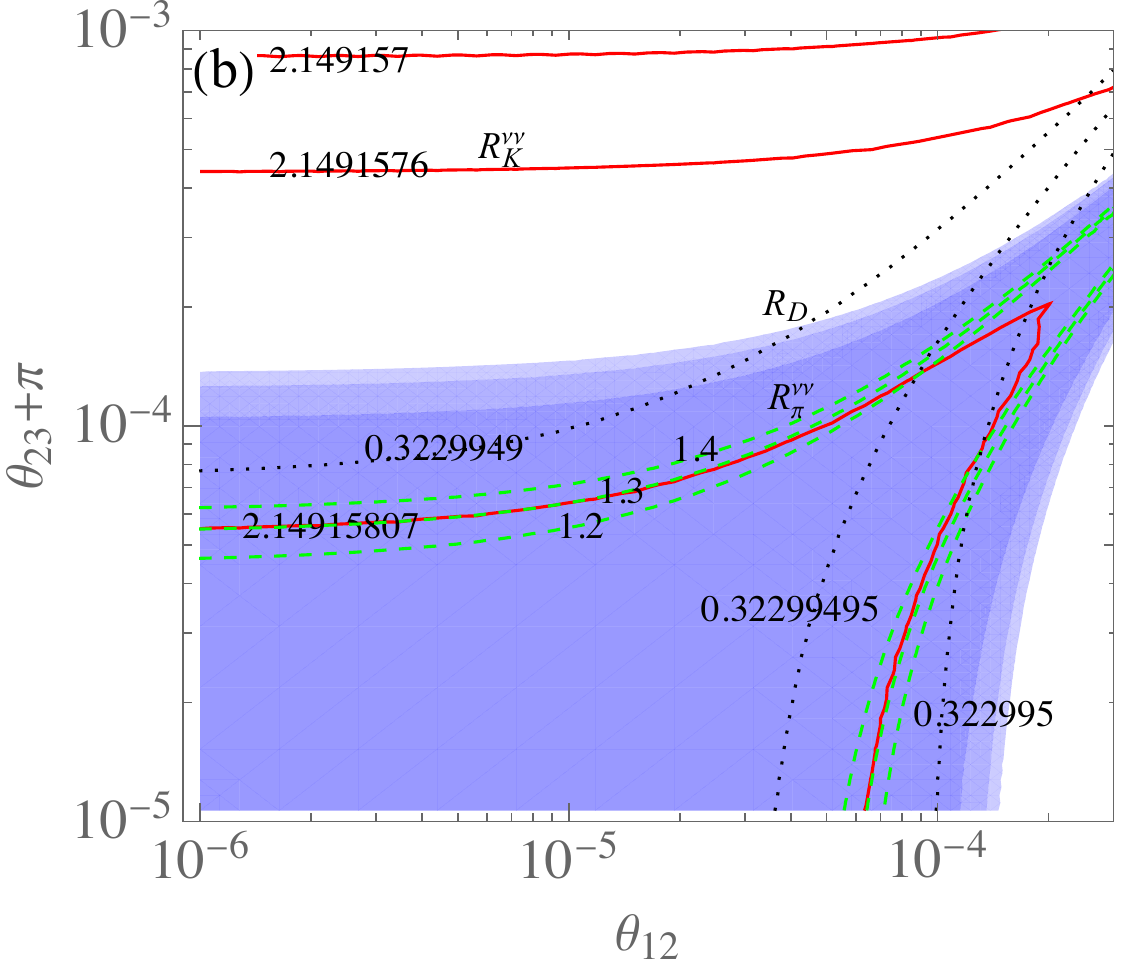}
    \vspace{10pt}
    \\
    \includegraphics[scale=0.42]{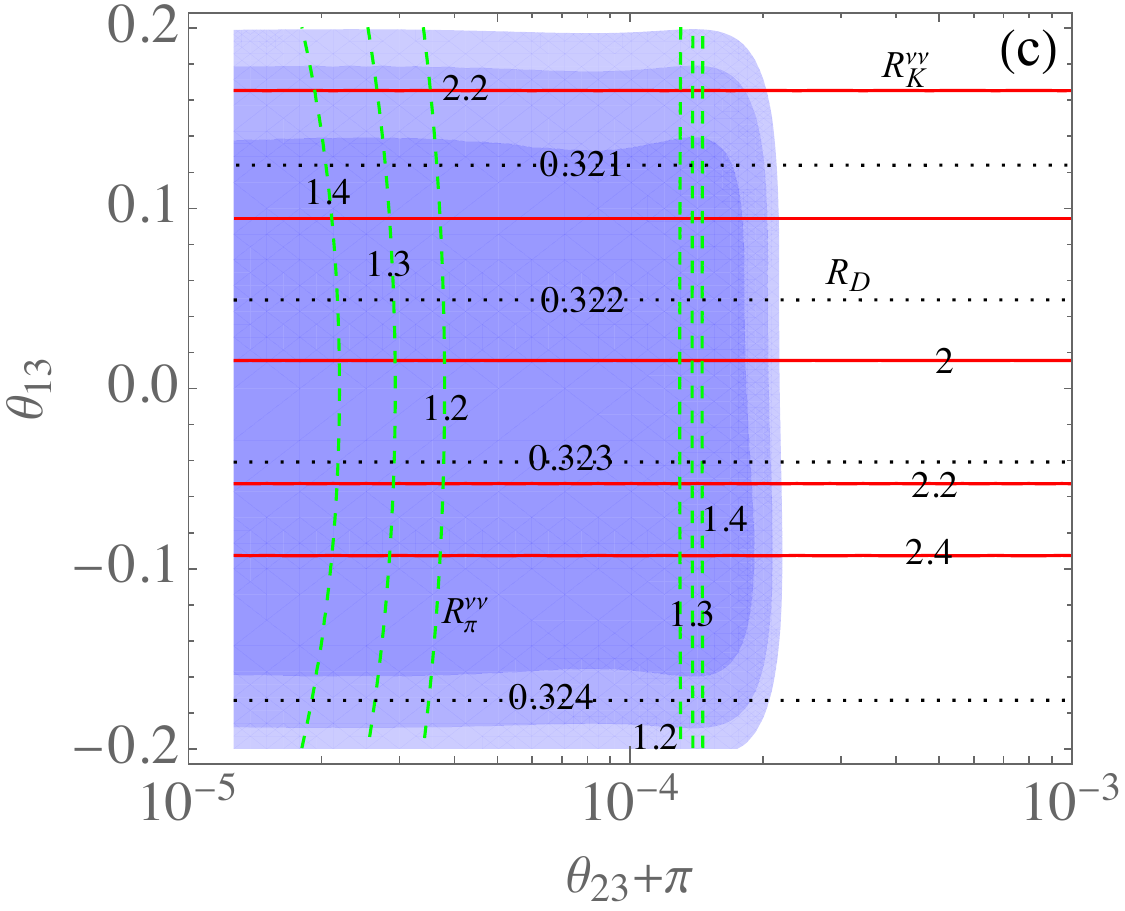}
    \includegraphics[scale=0.4]{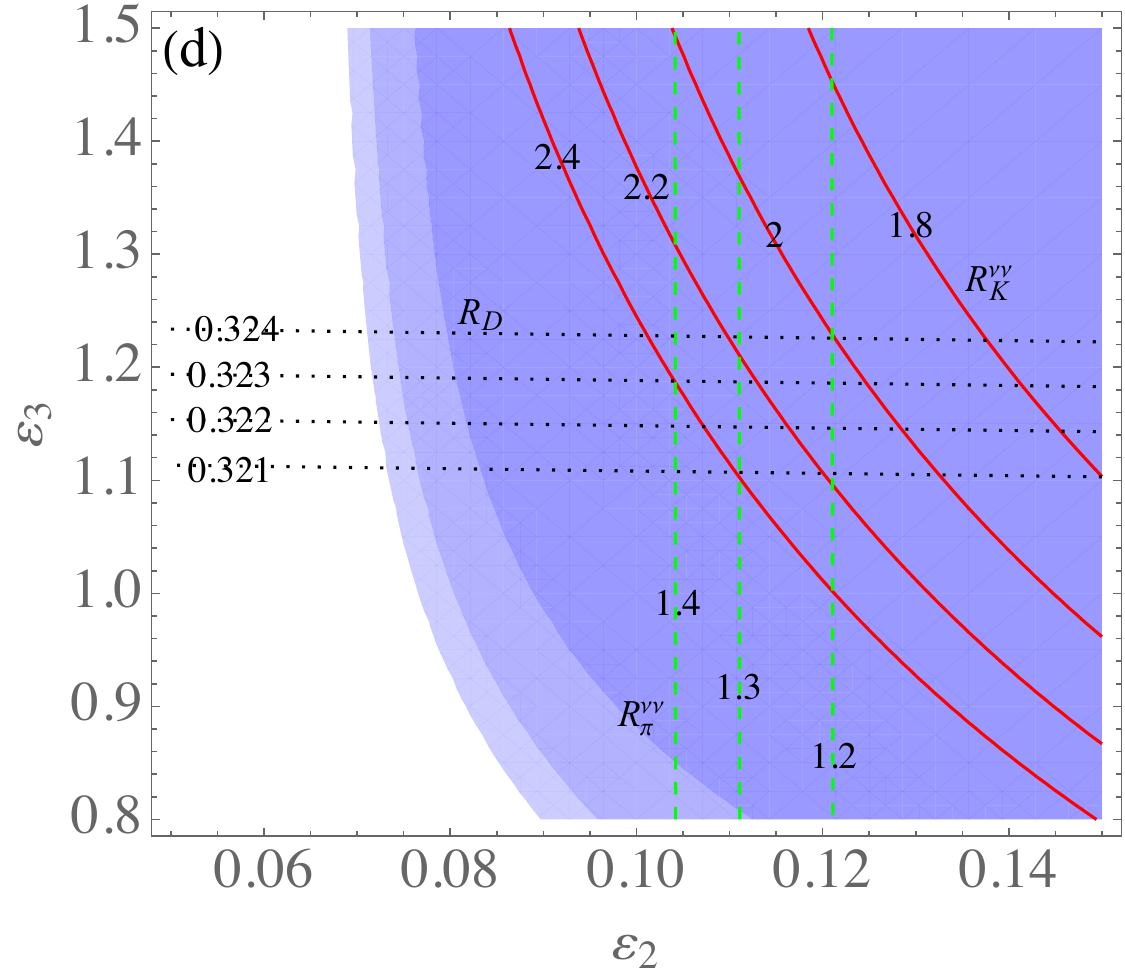}
    \caption{ Contours of $R^{\nu\nu}_K$ (red solid), $R^{\nu\nu}_{\pi^+}$ (green dashed), and $R_D$ (dotted) as functions of the selected two parameters. For comparison, the CLs are also shown in the plots. }
    \label{fig:obs_chi2}
\end{figure}

To investigate the influence of parameters on $R^{\nu\nu}_K$, $R^{\nu\nu}_{\pi^+}$, and $R_D$, we present contour plots of these observables in the plane of the two selected parameters in Fig.~\ref{fig:obs_chi2}, following the same parameter choices as in Fig.~\ref{fig:CL}.  For ease of comparison, regions of the $68.27\%$, $95\%$, and $99 \%$ CLs are superimposed in the plots.  As shown in Fig.~\ref{fig:obs_chi2}(a)-(c), $R^{\nu\nu}_K$ and $R_D$ are insensitive to $\theta_{12}$ and $\theta_{23}$.  This insensitivity is particularly evident in Fig.~\ref{fig:obs_chi2}(b), where their values remain nearly constant at  $2.15$ and $0.323$, respectively.  Since $\varepsilon_{2,3}$ are fixed at their optimized values, $R^{\nu\nu}_K$ and $R_D$ show only minor variations around $2.0$ and $0.323$ in Figs.~\ref{fig:obs_chi2}(a) and (c).

As shown in the contour plots in Fig.~\ref{fig:obs_chi2}(d), $R^{\nu\nu}_K$ exhibits a wider range of possible values and has a stronger dependence on $\varepsilon_{2,3}$.  In contrast, $R^{\nu\nu}_{\pi^+}$ is insensitive to $\theta_{13}$ and $\epsilon_3$, with its variations confined to very narrow regions of $\theta_{12}$ and $\theta_{23}$.  Additionally, Fig.~\ref{fig:obs_chi2}(d) shows that $R^{\nu\nu}_{\pi^+}$ is sensitive to $\varepsilon_2$.  From these results, we see clearly that both $R^{\nu\nu}_K$ and $R^{\nu\nu}_{\pi^+}$ can be significantly enhanced.  Although the enhancement of $R_D$ in the model is less pronounced, it still remains within the $1\sigma$ uncertainties of the current average value.

To assess the impact of the Yukawa couplings to the right-handed neutrinos on the processes $B\to K \nu \bar\nu$ and $K^+\to \pi^+ \nu \bar\nu$, we numerically evaluate the relevant parameter values.  According to Eqs.~(\ref{eq:Cij_B}) and (\ref{eq:Xij_K}), the left-handed and right-handed neutrino-related couplings in the model are defined as: 
 \begin{equation}
 Y^{q,ij}_{LL} \equiv (\tilde{\xi}_2)_{qj} (\tilde{\xi}_2)^\dagger_{i2},~ 
 Y^{q,ij}_{RR} \equiv (\tilde{\xi}_3)_{qj} (\tilde{\xi}_3)^\dagger_{i2}\,,
 \end{equation}
where $q=3(1)$ corresponds to the $B(K)$-meson decays.  Using the optimized parameters given in Eq.~(\ref{eq:optimal_values}), the numerical values of $Y^{ij}_{LL}$ and $Y^{ij}_{RR}$ are obtained as:
 \begin{equation}
 \begin{split}
 Y^{q=1}_{LL}  &=
\left(
\begin{array}{ccc}
 0 &  0  &  0  \\
0  &  1.27 & -1.89  \\
 0 &   -1.24 &  1.84
\end{array}
\right) \times 10^{-6},~~
Y^{q=1}_{RR}  =
\left(
\begin{array}{ccc}
0 & 0  & 0  \\
 0 &  1.97 & -0.11   \\
 0  &   -0.07 &  0.004
\end{array}
\right) \times 10^{-3}, 
\\
Y^{q=3}_{LL}  & =
\left(
\begin{array}{ccc}
 0 & 0  & 0  \\
 0 &  0.04 & 5.99  \\
 0 &  -0.04 &  -5.84
\end{array}
\right)\times 10^{-3},~~
Y^{q=3}_{RR}  =
\left(
\begin{array}{ccc}
 0 & 0  & 0  \\
 0 &  0.057 & 0.350   \\
  0 &   -0.002 &  -0.013 
\end{array}
\right)~.
\end{split}
 \end{equation} 
The zero entries in the first row and first column of $Y^{q}_{LL, RR}$ reflect our assumption of the vanishing mass of the lightest neutrino.  It is evident that numerically $Y^{q}_{RR}\gg Y^q_{LL}$, and $Y^{q=3}_{RR}\gg Y^{q=1}_{RR}$.  As discussed earlier, besides $\xi_4$, LFV processes are also affected by $\xi_1$.  From Eq.~(\ref{eq:xi_1}), $\xi_1$ is related to $\xi_2$ through the CKM and PMNS matrices.  Consequently, $\xi_2$ is constrained by LFV processes, such as $\mu$-$e$ conversion, $\mu\to e \gamma$, and $\tau\to \ell \gamma$.  Moreover, from Eq.~(\ref{eq:para_1}), we have $\tilde{\xi}_2 \propto \epsilon^\dagger$ and $\tilde{\xi}_3 \propto \epsilon^{-1}$.  Thus, when $\epsilon$ in $\tilde{\xi}_2$ is tightly constrained to smaller values, $\tilde{\xi}_3$ becomes enhanced due to its inverse dependence on $\epsilon$.  As a result, $Y^{q}_{LL} \ll Y^{q}_{RR}$, making the right-handed neutrino couplings the dominant contributors to the $B\to K \nu \bar\nu$ and $K^+\to \pi^+ \nu \bar\nu$ processes in this model.

\begin{figure}[tbph]
    \centering
    \includegraphics[scale=0.4]{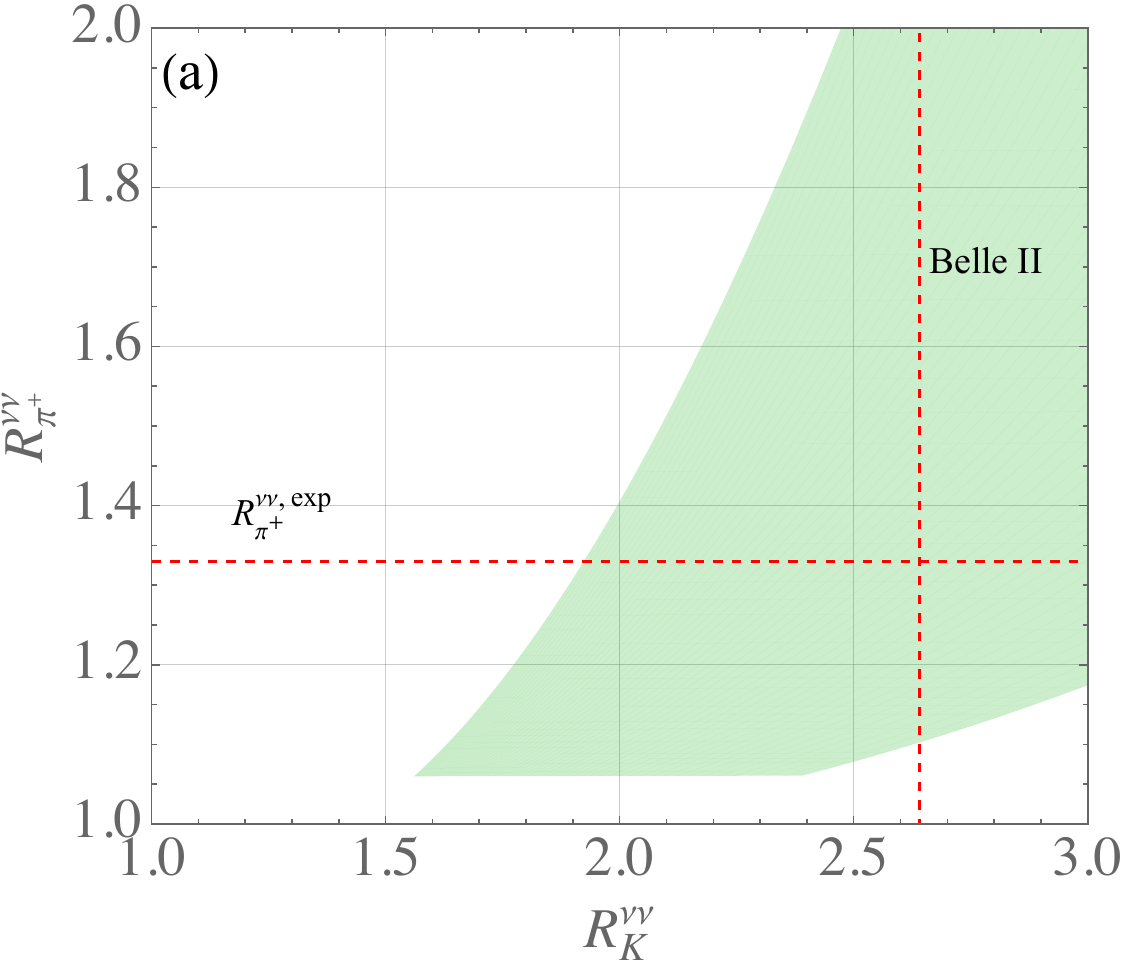}
    \hspace{0.4mm}
    \includegraphics[scale=0.41]{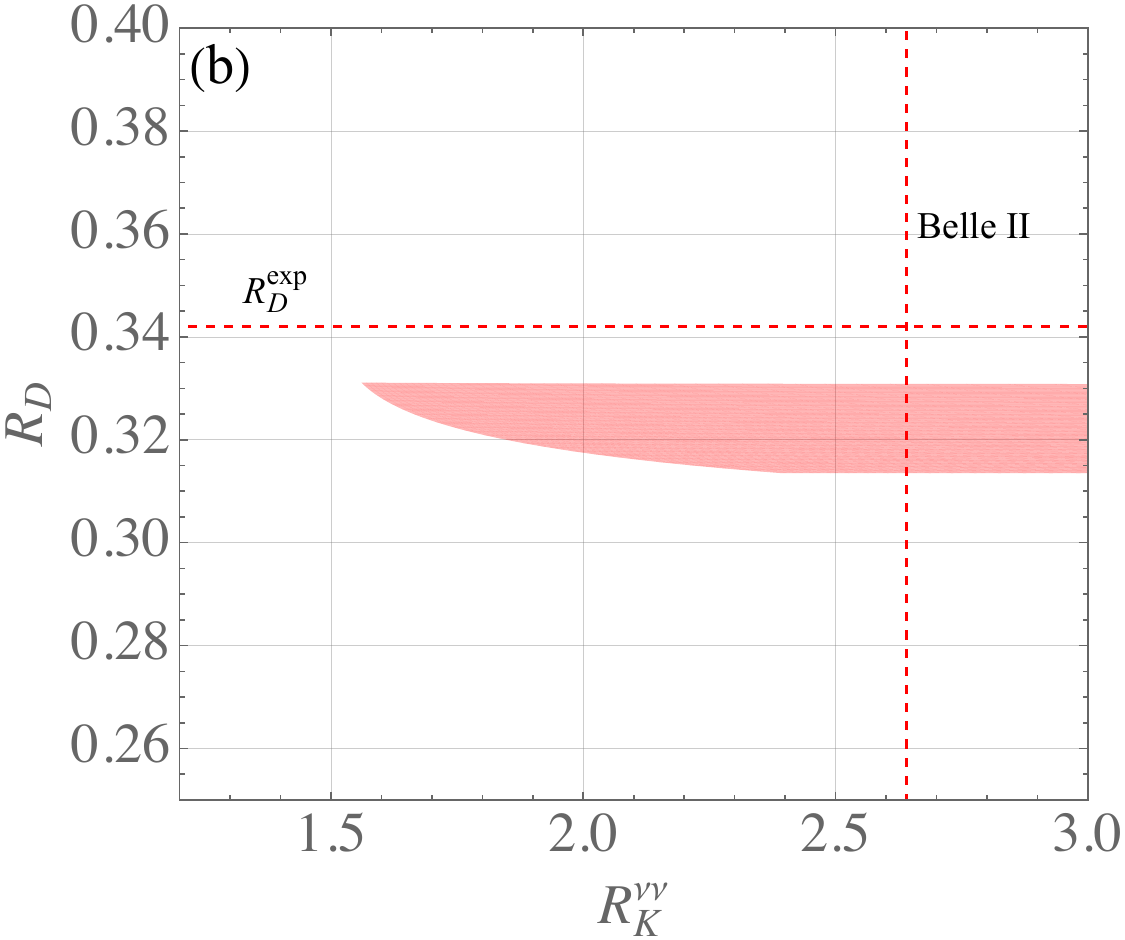} 
    \\
    \includegraphics[scale=0.4]{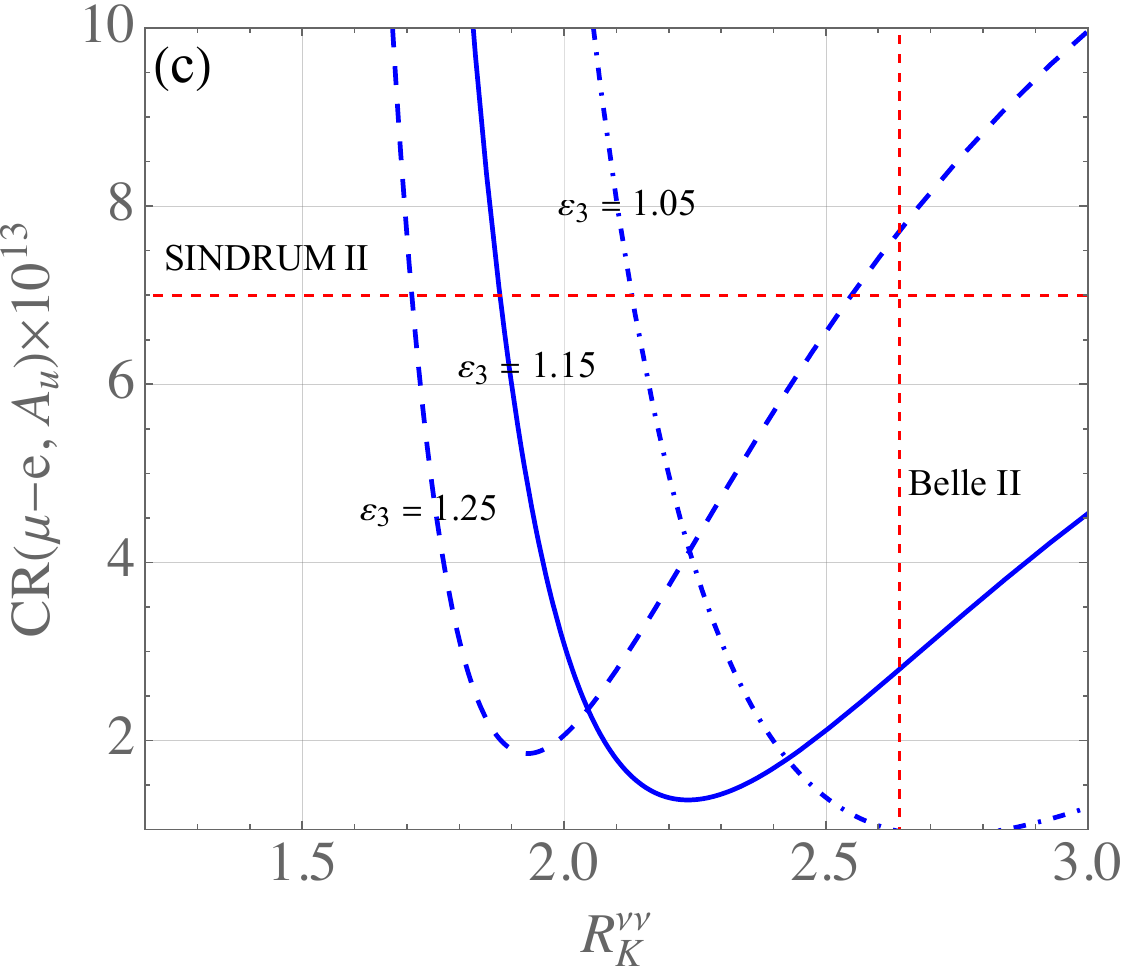}
    \hspace{1.2mm}
    \includegraphics[scale=0.4]{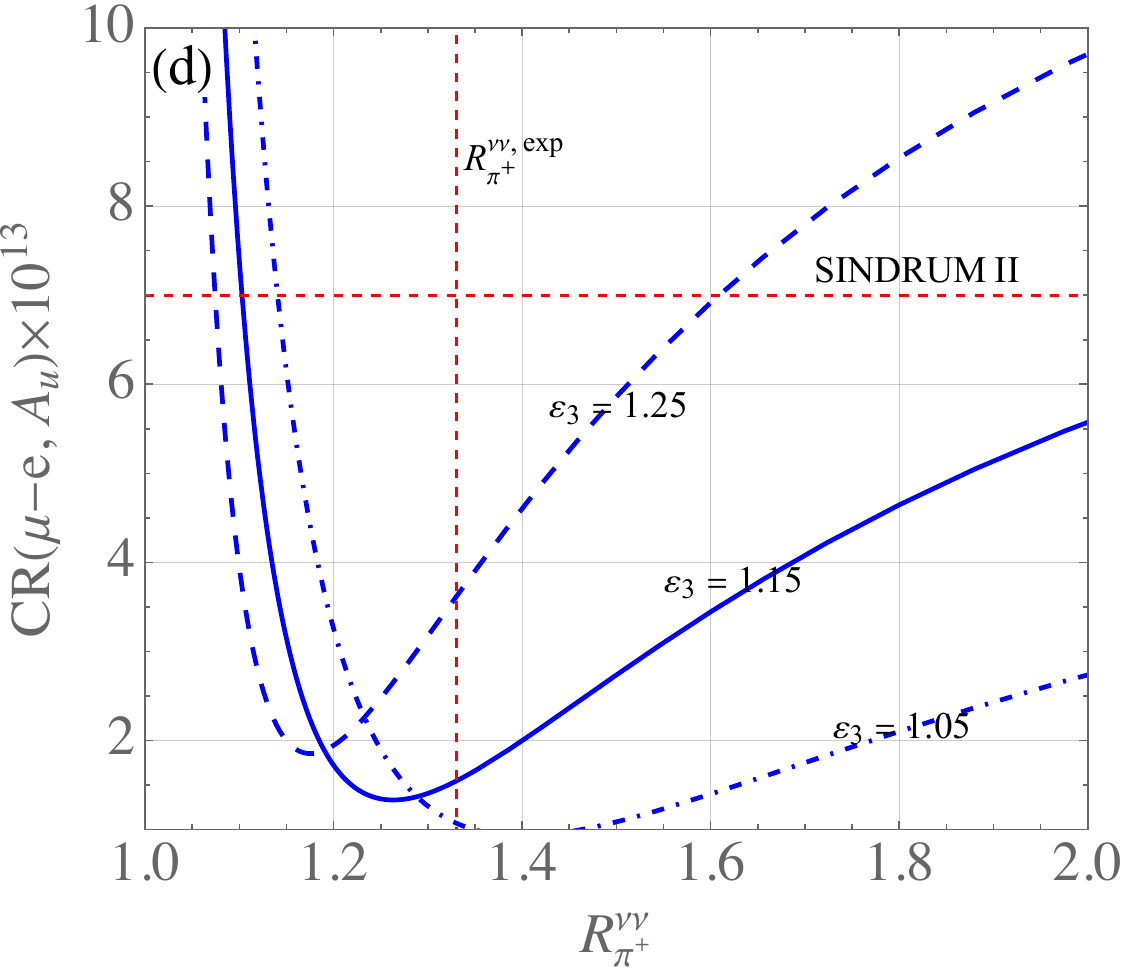}
    \caption{Correlation plots for (a) $R^{\nu\nu}_K$ vs. $R^{\nu\nu}_{\pi^+}$, (b) $R^{\nu\nu}_K$ vs. $R_D$, (c)  $CR(\mu-e, {\rm Au})$ vs. $R^{\nu\nu}_K$, and (d) $CR(\mu-e, {\rm Au})$ vs. $R^{\nu\nu}_{\pi^+}$.  The parameter ranges $\varepsilon_2\in (0.05,0.15)$ and $\varepsilon_3\in (0.8, 1.5)$ are used in plots (a) and (b), while $\varepsilon_2\in (0.05,0.15)$ is taken in plots (c) and (d).  Vertical and horizontal lines indicate the central values of the current experimental data, except that in plots (c) and (d) the horizontal lines represent the upper limit from SINDRUM II. }
    \label{fig:obs_obs}
\end{figure}

After illustrating the dependence of observables on the free parameters, it is insightful to examine their correlations.  Fig.~\ref{fig:obs_obs}(a) and (b) show the correlations of $R^{\nu\nu}_{K}$ with both $R^{\nu\nu}_{\pi^+}$ and $R_D$, respectively, where we vary the parameters $\varepsilon_2\in (0.05,0.15)$ and $\varepsilon_3\in (0.8, 1.5)$.  The remaining parameters are fixed at their optimized values, determined using the pattern $(\epsilon_d, \eta_d)$.  The horizontal and vertical dashed lines represent the central values of the current experimental measurements.  Fig.~\ref{fig:obs_obs}(c) and (d) show the correlations between $CR(\mu-e, \rm{Au})$ and $R^{\nu\nu}_K$ as well as $R^{\nu\nu}_{\pi^+}$, respectively, with $\varepsilon_2\in (0.05,1.5)$.  The dashed, solid, and dot-dashed curves correspond to $\varepsilon_3=1.25$, $1.15$, and $1.05$, respectively.  The vertical dashed lines indicate the central values of the current data, while the horizontal dashed lines are the upper limit of $CR(\mu-e, {\rm Au})$ measured by SINDRUM II~\cite{SINDRUMII:2006dvw}.

\subsection{Implication of future $\mu$-$e$ conversion sensitivity }

Due to the distinctive couplings of the LQs to leptons and quarks, a muon captured by a nucleus can undergo conversion to an electron via the process $\mu N \to e N$ at tree level, as demonstrated in this study.  Although the current experimental upper limit for this process is only of ${\cal O}(10^{-13})$, experiments such as Mu2e at Fermilab~\cite{Bernstein:2019fyh} and COMET~\cite{Moritsu:2022lem} at J-PARC, both using aluminum as the target nucleus, are expected to achieve sensitivities of ${\cal O}(10^{-17})$, and the sensitivities will be pushed to ${\cal O}(10^{-18})$ in the Mu2e-II and PRISM experiments~\cite{Lee:2022moh}.  It is thus natural to ask if the $\mu$-$e$ conversion is constrained or observed at the ${\cal O}(10^{-17})$ level, whether $B\to K\nu\bar\nu$ and $K^+\to \pi^+ \nu \bar\nu$ will be significantly enhanced in this model.

Since LFV processes are sensitive to $\zeta_{11, 21, 22}$, we demand that $CR(\mu-e, \rm{Al}) \sim {\cal O}(10^{-17})$ to determine their allowed ranges as $\zeta_{11,21}\in (10^{-9}, 10^{-7})$ and $\zeta_{22} \in (10^{-8}, 10^{-6})$.  Under these conditions, the key predictions are summarized as follows:
 \begin{equation}
 \begin{split}
 \chi^2_{\rm min} & =9.35, ~~ 
 {\cal B}(\mu\to e\gamma)=4.53\times 10^{-20}, ~~
 CR(\mu-e, \rm{Al})=0.95\times 10^{-17}, 
 \\
 {\cal B}(B\to K \nu \bar\nu)&=1.10\times 10^{-5},~~  
 {\cal B}(K^+\to \pi^+ \nu \bar\nu)=1.17\times 10^{-10},~~
 R_D=0.316\,.
 \end{split}
 \end{equation}
The corresponding optimized parameter values are:
 \begin{equation}
 \begin{split}
 \theta_{12}& =3.3\times 10^{-3},~~ 
 \theta_{13}= 0.0496,~~
 \theta_{23}=-3.138,~~
 \zeta_{11}=7.86\times 10^{-8},
 \\
 \zeta_{21}&=8.79\times 10^{-9},~~
 \zeta_{22}=6.07\times 10^{-7},~~
 \varepsilon_2=-0.117,~~
 \varepsilon_3=0.886\,.
 \end{split}
 \end{equation}
These results clearly indicate that even when $CR(\mu-e, \rm{Al})$ reaches the ${\cal O}(10^{-17})$ level, the enhancements in the $B\to K \nu \bar\nu$ and $K^+\to \pi^+ \nu \bar\nu$ processes can still be pronounced.  Additionally, the resulting value of $R_D$ is within the lower $1\sigma$ bound of current experimental data.

\section{Summary} \label{sec:summary}

We have proposed a model for generating Dirac-type neutrino masses radiatively.  As a minimal extension of the SM, the model introduces two scalar $S_1$ leptoquarks (LQs) in addition to three right-handed neutrinos.  The model also features a global $U(1)_X$ symmetry that plays a crucial role in suppressing tree-level Dirac mass contributions from the Brout-Englert-Higgs mechanism and in preventing leptoquarks from coupling directly to quark pairs.  The smallness of the neutrino mass arises from the soft breaking of the $U(1)_X$ symmetry in the scalar potential, which also induces mixing between the two LQs.

A non-Casas-Ibarra-type parametrization for Yukawa couplings that relates them to the neutrino masses and PMNS matrix is proposed to reduce the number of free parameters in the model.  Our $\chi^2$ analyses suggest that a vanishing mass for the lightest neutrino is preferred, with the normal mass ordering scenario generally yielding lower $\chi^2_{\rm min}$ values than the inverted mass ordering scenario.  Accordingly, we focus on the normal ordering scenario with $m_{\nu_1}=0$ and employ a $\chi^2$ fit to nine observables to determine the optimized values of eight model parameters.

When the current experimental upper bounds on LFV processes are imposed, the left-handed neutrino Yukawa couplings are stringently constrained. In contrast, the right-handed neutrino Yukawa couplings become the dominant contributors to enhance the $B\to K^{(*)} \nu \bar\nu$ and $K\to \pi \nu \bar\nu$ decays.  We find that the branching ratio, relative to the SM prediction, can be enhanced up to a factor of $2$ for $B\to K^{(*)} \nu \bar\nu$ and $1.2$ for $K^+\to \pi^+ \nu \bar\nu$.  Since our analysis employs a simplified non-Casas-Ibarra parametrization without introducing a new CP-violating phase, the CP-violating process $K_L\to \pi \nu \bar\nu$ remains unaffected.  Interestingly, within this framework, the ratio $R_D$ can be enhanced to fall within the $1\sigma$ range of the current experimental data.

Finally, we examine the impact of future experimental sensitivity in the $\mu$-$e$ conversion, particularly when it reaches the ${\cal O}(10^{-17})$ level, as projected by Mu2e and COMET.  Under these conditions, the branching ratios of $B\to K^{(*)} \nu \bar\nu$ and $K^+\to \pi^+ \nu \bar \nu$ can still have significant enhancements, while $R_D$ still falls within the $1\sigma$ lower bound of the current data.

\bibliographystyle{apsrev4-1}
\bibliography{DiracMass}

\end{document}